\documentclass{article}

\usepackage[margin=1in,footskip=0.25in]{geometry} \usepackage[dvipsnames,table]{xcolor}             \DeclareMathAlphabet{\mathcal}{OMS}{cmsy}{m}{n}

\usepackage[sort&compress,square,comma,numbers]{natbib} \usepackage[pagebackref]{hyperref}                \usepackage{url}                                  \usepackage{breakcites}                           

\usepackage[nosumlimits]{amsmath} \usepackage{amsthm}               \usepackage{amssymb}              \usepackage{amsfonts}             \usepackage{bm}                   \usepackage{bbm}                  \usepackage{mathtools}            \usepackage{nicefrac}             \usepackage{stmaryrd}             \usepackage{commath}              

\usepackage{fix-cm}         \usepackage{graphicx}       \usepackage{tikz}           \usetikzlibrary{arrows.meta,calc,decorations.pathreplacing,patterns}
\usepackage{float}          \usepackage{caption}        \usepackage{subcaption}     \usepackage{booktabs}       \usepackage[most]{tcolorbox}  

\usepackage{algorithm}      \usepackage{algpseudocode}  \usepackage{algorithmicx}   

\usepackage{enumitem}       \usepackage{xspace}         \usepackage{relsize}        \usepackage{titlesec}       \usepackage[normalem]{ulem} \usepackage{contour}        \usepackage{cancel}         \usepackage{todonotes}      \usepackage[capitalize,nameinlink,noabbrev]{cleveref} \usepackage{crossreftools}

\hypersetup{
    colorlinks=true,        linkcolor=PineGreen,    citecolor=NavyBlue,     filecolor=Purple,       urlcolor=Purple         }

\renewcommand*{\backref}[1]{}
\renewcommand*{\backrefalt}[4]{\ifcase #1 (No citations.)\or
        (Cited on page~#2.)\else
        (Cited on pages~#2.)\fi
}

\titlespacing*{\section}{0pt}{12pt}{5pt}
\titlespacing*{\subsection}{0pt}{11pt}{5pt}
\titlespacing*{\subsubsection}{0pt}{11pt}{5pt}
\titlespacing*{\paragraph}{0pt}{6pt}{1em}
\newcommand{\polyfont}{\fontsize{9.3}{11.2}\selectfont}
\newcommand{\smallfont}{\fontsize{8.4}{10.3}\selectfont}
\newcommand{\titlefont}{\fontsize{10}{12}\selectfont\bfseries}
\tikzset{
  every node/.style={font=\polyfont,inner sep=2pt},
  head/.style={font=\titlefont,anchor=west},
  minor/.style={font=\smallfont},
  edge/.style={-{Stealth[length=1.7mm,width=1.3mm]},line width=.55pt},
  axis/.style={-{Stealth[length=1.5mm,width=1.1mm]},line width=.5pt},
  rule/.style={line width=.45pt,black!50},
  guide/.style={densely dashed,line width=.45pt,black!55},
  curve/.style={line width=.95pt,line cap=round,line join=round},
  panel/.style={draw=black!60,line width=.5pt,rounded corners=1.3pt},
  boxtext/.style={align=center,text width=4.8cm},
}

\floatstyle{plaintop}
\restylefloat{table}

\algrenewcommand\alglinenumber[1]{\sf\scriptsize\color{NavyBlue}{#1}} 

\let\oldthebibliography\thebibliography
\renewcommand{\thebibliography}[1]{\oldthebibliography{#1}\setlength{\itemsep}{0pt}\setlength{\parskip}{0pt}}

\newtheorem{theorem}{Theorem}[section]

\newtheorem{lemma}[theorem]{Lemma}
\newtheorem{fact}[theorem]{Fact}

\newtheorem{setting}[theorem]{Setting}

\newtheorem{importedtheorem}[theorem]{Imported Theorem}
\newtheorem{importedlemma}[theorem]{Imported Lemma}

\theoremstyle{definition}
\newtheorem{definition}[theorem]{Definition}

\theoremstyle{remark}

\Crefname{setting}{Setting}{Settings}
\Crefname{problem}{Problem}{Problems}
\Crefname{equation}{Equation}{Equations}
\Crefname{importedtheorem}{Imported Theorem}{Imported Theorems}
\Crefname{importedlemma}{Imported Lemma}{Imported Lemmas}

\theoremstyle{theorem}
\makeatletter
\newtheorem*{rep@theorem}{\rep@title}
\newcommand{\newreptheorem}[2]{\newenvironment{rep#1}[1]{\def\rep@title{\Cref{##1}, restated}\begin{rep@theorem}}{\end{rep@theorem}}}
\makeatother

\newreptheorem{theorem}{Theorem}
\newreptheorem{lemma}{Lemma}
\newreptheorem{definition}{Definition}
\newreptheorem{corollary}{Corollary}

\contourlength{0.8pt}
\newcommand{\unline}[1]{\uline{\phantom{#1}}\llap{\contour{white}{#1}}}

\makeatletter
\def\hlinewd#1{\noalign{\ifnum0=`}\fi\hrule \@height #1 \futurelet
	\reserved@a\@xhline}
\makeatother

\DeclareMathOperator*{\E}{\mathbb{E}}
\DeclareMathOperator{\tr}{tr}

\DeclareMathOperator{\Span}{span}
\DeclareMathOperator{\cond}{cond}

\newcommand{\defeq}{\ensuremath{\;{\vcentcolon=}\;}\xspace}
  \newcommand{\eps}[0]{\varepsilon} \let\epsilon\eps

\let\norm\relax
\newcommand{\norm}[1]{\|#1\|}

\newcommand{\mat}[1]{\bm{#1}}    \renewcommand{\vec}[1]{\bm{#1}}    

 \newcommand{\bmat}[1]{\begin{bmatrix} #1 \end{bmatrix}}

\newcommand{\mA}{\ensuremath{\mat{A}}\xspace}
\newcommand{\mB}{\ensuremath{\mat{B}}\xspace}
\newcommand{\mC}{\ensuremath{\mat{C}}\xspace}

\newcommand{\mF}{\ensuremath{\mat{F}}\xspace}
\newcommand{\mG}{\ensuremath{\mat{G}}\xspace}
\newcommand{\mH}{\ensuremath{\mat{H}}\xspace}
\newcommand{\mI}{\ensuremath{\mathbf{I}}\xspace}

\newcommand{\mK}{\ensuremath{\mat{K}}\xspace}
\newcommand{\mL}{\ensuremath{\mat{L}}\xspace}

\newcommand{\mP}{\ensuremath{\mat{P}}\xspace}
\newcommand{\mQ}{\ensuremath{\mat{Q}}\xspace}
\newcommand{\mR}{\ensuremath{\mat{R}}\xspace}
\newcommand{\mS}{\ensuremath{\mat{S}}\xspace}
\newcommand{\mT}{\ensuremath{\mat{T}}\xspace}
\newcommand{\mU}{\ensuremath{\mat{U}}\xspace}
\newcommand{\mV}{\ensuremath{\mat{V}}\xspace}
\newcommand{\mW}{\ensuremath{\mat{W}}\xspace}

\newcommand{\mZ}{\ensuremath{\mat{Z}}\xspace}

\newcommand{\mDelta}{\ensuremath{\mat{\Delta}}\xspace}
\newcommand{\mSigma}{\ensuremath{\mat{\Sigma}}\xspace}

\newcommand{\va}{\ensuremath{\vec{a}}\xspace}
\newcommand{\vb}{\ensuremath{\vec{b}}\xspace}
\newcommand{\vc}{\ensuremath{\vec{c}}\xspace}

\newcommand{\ve}{\ensuremath{\mathbf{e}}\xspace}

\newcommand{\vg}{\ensuremath{\vec{g}}\xspace}
\newcommand{\vh}{\ensuremath{\vec{h}}\xspace}

\newcommand{\vl}{\ensuremath{\vec{\ell}}\xspace}
\newcommand{\vm}{\ensuremath{\vec{m}}\xspace}

\newcommand{\vq}{\ensuremath{\vec{q}}\xspace}
\newcommand{\vr}{\ensuremath{\vec{r}}\xspace}

\newcommand{\vu}{\ensuremath{\vec{u}}\xspace}
\newcommand{\vv}{\ensuremath{\vec{v}}\xspace}
\newcommand{\vw}{\ensuremath{\vec{w}}\xspace}
\newcommand{\vx}{\ensuremath{\vec{x}}\xspace}
\newcommand{\vy}{\ensuremath{\vec{y}}\xspace}
\newcommand{\vz}{\ensuremath{\vec{z}}\xspace}

\newcommand{\cA}{\ensuremath{{\mathcal A}}\xspace}

\newcommand{\cD}{\ensuremath{{\mathcal D}}\xspace}
\newcommand{\cE}{\ensuremath{{\mathcal E}}\xspace}
\newcommand{\cF}{\ensuremath{{\mathcal F}}\xspace}

\newcommand{\cI}{\ensuremath{{\mathcal I}}\xspace}

\newcommand{\cK}{\ensuremath{{\mathcal K}}\xspace}

\newcommand{\cM}{\ensuremath{{\mathcal M}}\xspace}
\newcommand{\cN}{\ensuremath{{\mathcal N}}\xspace}
\newcommand{\cO}{\ensuremath{{\mathcal O}}\xspace}

\newcommand{\cR}{\ensuremath{{\mathcal R}}\xspace}
\newcommand{\cS}{\ensuremath{{\mathcal S}}\xspace}
\newcommand{\cT}{\ensuremath{{\mathcal T}}\xspace}

\newcommand{\cW}{\ensuremath{{\mathcal W}}\xspace}

\newcommand{\bbP}{\ensuremath{{\mathbb P}}\xspace}

\newcommand{\bbR}{\ensuremath{{\mathbb R}}\xspace}
\newcommand{\bbS}{\ensuremath{{\mathbb S}}\xspace}

\newcommand{\rC}{\ensuremath{\mathrm{C}}\xspace}

\usepackage{xcolor}
\usepackage{multirow}

\title{From smoothed analysis to backward error:\\
The power of adaptive perturbations
}

\author{\begin{tabular}[t]{c@{\hspace{4em}}c}
    Micha{\l} Derezi{\'n}ski & Michael W. Mahoney \\
    University of Michigan & UC Berkeley, ICSI, LBNL \\
    \texttt{derezin@umich.edu} & \texttt{mmahoney@stat.berkeley.edu} \\[2.5ex]
    Raphael A. Meyer & Christopher Musco \\
    UC Berkeley, ICSI, LBNL & New York University \\
    \texttt{ram900@berkeley.edu} & \texttt{cmusco@nyu.edu}
  \end{tabular}}
\date{\today}

\begin{document}

\maketitle

\begin{abstract}
Smoothed analysis has become a central tool in theoretical computer science, often allowing for significant improvements to worst-case running times if we settle for a solution to a small \emph{random} perturbation of the problem instance.
This is 
closely related to the classical concept of \emph{backward error} in numerical analysis, which allows an algorithm to return a solution to an \emph{arbitrary} perturbation of the problem.

Despite the qualitative similarity of these concepts, we show that they can exhibit large quantitative differences.
Notably, we prove that, while a linear system with any $n \times n$ matrix $\mA$ can be solved to backward error $\eps$ using $\cO(\log(n) / \eps)$ matrix-vector products with $\mA$, any algorithm that solves a random $\eps$-sized perturbation of $\mA$ requires $\Omega(n)$ matrix-vector products. This means that a useful perturbation for solving a linear system in backward error must be \emph{adaptive} to the input. Interestingly, however, randomness is still essential: we show that any deterministic method that attains small backward error for all linear systems provably requires $\Omega(n)$ matrix-vector products, resulting in the first separation between randomized and deterministic algorithms in a natural computational model for solving linear systems. Finally, we show that matrix structure can significantly affect backward error complexity. For example, we show a deterministic $\cO(1/\eps)$ matrix-vector product algorithm for symmetric linear systems.
\end{abstract}

\section{Introduction}

The \emph{smoothed analysis} framework of Spielman and Teng has proven to be a powerful tool for analyzing algorithms whose worst-case performance is brittle \cite{spielman2009smoothed,roughgarden2021beyond}. The idea is to prove that an algorithm solves a given problem efficiently on any \emph{randomly perturbed} input, with high probability. 

Smoothed analysis results can be interpreted in a few different ways.
One interpretation is that any real world data inevitably arrives perturbed due to measurement errors, roundoff errors, etc., so smoothed analysis captures performance on ``realistic'' inputs \cite{spielman2004smoothed,sankar2006smoothed}.
Another interpretation is that smoothed analysis gives a way to design algorithms that have small \emph{backward error}, meaning that they return a solution to a nearby problem instance.
In particular, we can randomly perturb our input before running an algorithm with low smoothed complexity. 
This second view has proven especially useful in recent years: smoothed analysis has been central in recent progress on $\widetilde\cO(n^3)$ time algorithms for matrix diagonalization \cite{banks2021gaussian,banks2023pseudospectral,Banks2024OverlapsEG,banks2025global,banks2025globalII,banks2025globalIII,shah2024sparse}, faster sparse linear system solvers \cite{nie2022matrix,peng2024solving},  Krylov  method analysis \cite{Meyer2023OnTU,Chen2026DoesBS,kressner2024randomized}, and more \cite{drineas2024stochastic,dexter2025stochastic}.

Crucially, smoothed analysis is not the \emph{only approach} to designing algorithms with small backward error.
Indeed, backward error has been studied for decades in numerical analysis and other fields, dating back to Turing, von Neumann, and Wilkinson, among others \cite{turing1948rounding,von1947numerical,wilkinson1960error}, and was motivated by the need to account for the noise from finite precision roundoff and other sources \cite{wilkinson1963rounding,rice1966theory,higham2002accuracy}. 
Thereby, a variety of numerical methods have been designed to minimize backward error without the use of random perturbations.

To form a more rigorous taxonomy, we can categorize backward error algorithms by the class of perturbations they permit.
Taking linear systems as a running example, for any input matrix $\mA$ and vector $\vb$, an algorithm for solving $\mA\vx = \vb$ returns an $\eps$ backward error solution \(\widetilde\vx\) if
\begin{equation}
    \label{eq:backward-err-defn}
    (\mA+\mDelta)\widetilde\vx=\vb,
    \quad \norm{\mDelta}_2\leq \eps\norm{\mA}_2=:\eta
    \quad\text{for}\quad
    \mDelta\,\sim\begin{cases}
     \cD(\eta)&\text{(oblivious perturbation)}\\
    \cD(\mA,\vb,\eta) &\text{(adaptive perturbation)}.
\end{cases}
\end{equation}
Here, $\cD$ indicates the distribution that generates the perturbation used by the algorithm. For algorithms in the smoothed analysis model, this distribution is \emph{oblivious} to the input up to scaling (first case).
On the other hand, the general backward error model allows any perturbation, random or deterministic,
which can be chosen \emph{adaptively} based on the input (second case).

This difference raises several natural research questions.
Are algorithms built on \emph{adaptive} perturbations actually more efficient than algorithms built on \emph{oblivious} ones?
Are there natural problems where achieving small backward error is possible with much less computation than smoothed analysis would suggest?
When is it important for adaptive perturbations to be randomized?

\section{Our Contributions}
\label{sec:contributions}

We initiate the study of the questions above in the context of linear system solving, which is one of the most important tasks in numerical analysis, and a problem for which smoothed analysis has been widely applied \cite{sankar2006smoothed,peng2024solving}. Specifically, we study system solving in the popular \emph{matrix-vector product} model of computation, where an algorithm is allowed to interact with the input matrix $\mA$ only through implicit queries of the form:
\begin{align*}
    \vx\mapsto \mA\vx\qquad\text{and}\qquad \vx\mapsto \mA^\top\vx.
\end{align*}
This computational model is rich enough to implement many of the most common algorithms in numerical linear algebra, including iterative Krylov methods and randomized matrix sketching methods \cite{saad2003iterative,greenbaum1997iterative,tropp2011structure,martinsson2020randomized,Woodruff_sketching_NOW}. At the same time, the model is restricted enough that it offers the possibility of proving strong lower bounds. Notably, there has been recent success in revealing fundamental algorithmic barriers through information-theoretic lower bounds in the matrix-vector product model, including for linear system solving \cite{nemirovskij1983problem,braverman2020gradient,chewi2024query,derezinski2026matrix,shamir2026simple}, matrix approximation \cite{simchowitz2018tight,bakshi2023krylov,chen2025near,amsel2026fixed,amsel2026query}, and a number of other tasks \cite{sun2021querying,meyer2021hutch}. 

Building on this work, we obtain a number of surprising conclusions on the difference between oblivious and adaptive perturbations in backward error analysis. We summarize our findings below:
\paragraph{1. An exponential gap between smoothed and backward error complexity.} 
    A recent algorithm of \cite{derezinski2026towards} shows how to produce an $\eps$ backward error solution to an $n\times n$ linear system using $\cO(\log(n)/\eps)$ matrix-vector products (an improvement over the trivial $n$ product algorithm) by relying on \emph{adaptive} perturbations. We prove that, on the other hand, smoothed analysis is unable to give \emph{any} non-trivial complexity (i.e., better than $\cO(n)$), regardless of the perturbation distribution $\cD$.
In other words, \emph{adaptive perturbations offer an exponential complexity improvement over what is possible with oblivious ones.}
\paragraph{2. Randomization reduces the complexity of solving linear systems.}
    Given that adaptive perturbations offer such significant gains over random oblivious ones, it is natural to ask whether we need randomness at all. We show that randomness is in fact necessary, proving that no non-trivial deterministic backward error guarantee is possible for general linear systems in the matrix-vector product model. Thus, neither adaptivity nor randomness is enough by itself: \emph{We need randomized \unline{and} adaptive perturbations}.

    \paragraph{3. Symmetry implies deterministic solvers beyond matrix multiplication.} 
In contrast to the result above, we show that merely imposing symmetry on the input matrix changes the story significantly: we obtain a deterministic algorithm that solves any symmetric linear system to small backward error faster than a direct solver equipped with fast matrix multiplication. This suggests that \emph{deterministic adaptive perturbations are nevertheless a powerful paradigm} for certain classes of problems.

\smallskip We discuss each of these three contributions in greater detail below. 

\subsection{
Contribution 1: Hardness for smoothed analysis
}
Our first result separating smoothed analysis and adaptive perturbation methods is inspired by a simple observation. In particular, 
it is possible to show that no matrix-vector product (matvec) algorithm can yield a non-trivial worst-case complexity guarantee for solving an $n\times n$ linear system $\mA\vx=\vb$ to small \emph{forward error}. Concretely, we need $\Omega(n)$ matvecs to obtain even a rough approximation $\widetilde\vx$ such that $\norm{\widetilde\vx - \vx}_2 \leq \frac12\norm{\vx}_2$, and this bound can be matched by the trivial procedure of reading the matrix $\mA$ column-by-column using $n$ matvecs and then solving $\mA\vx=\vb$ using a direct method like Gaussian elimination \cite{chewi2024query,derezinski2026matrix}.

Indeed, this $\Omega(n)$ lower bound can be proven by simply taking $\mA$ to be a random Gaussian matrix. As is well-known, $\mA$ will have singular value condition number $\Theta(n)$ with high probability \cite{edelman1988eigenvalues}, and thus standard iterative methods implementable in the matvec model will require $\Omega(n)$ steps to converge. 
In the original formulation of smoothed analysis, Spielman and Teng propose perturbing the problem input with a \emph{Gaussian} perturbation, $\mDelta$ \cite{spielman2009smoothed}. But if \mA itself is a Gaussian matrix, we observe that \(\mA+\mDelta\) is as well, so the exact same $\Omega(n)$ lower bound applies! Of course, to rule out better smoothed analysis methods, we need to show that \emph{no oblivious perturbation suffices}, not just a Gaussian one. But this observation is our first hint that it might not be possible to match the $\cO(\log (n)/\eps)$ matvec result of \cite{derezinski2026towards}, which is based on an adaptive perturbation, with an oblivious perturbation method.

Indeed, we are able to formalize this intuition, showing that an $n \times n$ linear system requires $\Omega(n)$ matvecs to solve in the smoothed analysis setting, even when allowing for \emph{arbitrary oblivious} perturbations:
 
\begin{theorem}[Smoothed analysis does not help]
    \label{thm:main-smoothed}
    There are universal constants $c,p\in(0,1)$ such that the following holds. Fix any distribution over random matrices \(\mDelta\in\bbR^{n \times n}\) such that \(\norm{\mDelta}_2 \leq c\). 
    For any matrix \(\mA\in\bbR^{n \times n}\) with \(\norm{\mA}_2 \le 1\), denote \(\widehat\mA = \mA+\mDelta\).
Any algorithm that takes a vector $\vb$, computes at most \(t\) matvecs with \(\widehat\mA\), and returns a vector \(\widetilde\vx\) such that 
    \(\norm{\widetilde\vx-\widehat\mA^{-1}\vb}_2\leq \frac1{20}\norm{\widehat\mA^{-1}\vb}_2\) with probability at least \(p\) must use \(t \gtrsim n\) matvecs in the worst case.
\end{theorem}
\cref{thm:main-smoothed} rules out the possibility of efficiently obtaining a small backward error solution to a linear system by perturbing $\mA$ and solving to small forward error, i.e., via the standard smoothed analysis approach. Indeed, this approach can be exponentially worse than the existing method of \cite{derezinski2026towards}, requiring $\Omega(n)$ instead of $\cO(\log n)$ matrix-vector products. 

We prove \cref{thm:main-smoothed} in \cref{sec:smoothed-lower-bound} by establishing hardness against an arbitrarily perturbed Gaussian system, combining several key ingredients: a ``hidden Gaussian theorem'' for capturing randomness unobserved by the algorithm (\cref{thm:hidden-gaussian}), a condition number tail estimate for a shifted Gaussian matrix (\cref{thm:gaussian-near-singularity}), and a reduction from approximating the Frobenius norm of an inverse (\cref{lem:gaussian-factor-two-estimation}).

\subsection{
Contribution 2: Hardness for deterministic algorithms
}
Given that oblivious random perturbations yield no non-trivial matvec guarantee for solving linear systems, it is natural to ask if randomization is even a necessary algorithmic tool for obtaining small backward error solutions to linear systems.
Indeed, despite its remarkable successes in other numerical tasks such as low-rank approximation \cite{cw-sparse}, trace estimation \cite{hutchinson1989stochastic,avron2011randomized}, and spectral density estimation \cite{chen2021analysis,braverman2022sublinear}, randomness has had limited impact on the complexity of solving general, unstructured linear systems \cite{derezinski2026matrix,kelner2026solving}. Indeed, in the common setting where $\mA$ is assumed to have bounded condition number, it is known that randomized matvec algorithms cannot even improve on the leading constant of the best deterministic methods for recovering a vector \(\widetilde\vx\) with small residual \(\norm{\mA\widetilde\vx-\vb}_2 \leq \eps\norm{\vb}_2\) \cite{shamir2026simple}. Is the value of randomness similarly limited when our goal is to obtain small backward error without any assumptions on $\mA$?

We show that, surprisingly, this is not the case: no deterministic algorithm can achieve small backward error in fewer than $\Omega(n)$ matvecs on all $n\times n$ linear systems, meaning that adaptive deterministic perturbations yield no better matvec complexity than the trivial algorithm. 
\begin{theorem}[Deterministic methods cannot achieve small backward error]
    \label{thm:main-deterministic}
    Any deterministic algorithm that takes a vector \vb, computes at most \(t\) matvecs with a matrix $\mA$, and returns a vector \(\widetilde\vx\) that solves \((\mA+\mDelta)\widetilde\vx=\vb\) such that $\norm{\mDelta}_2\leq \frac12\norm{\mA}_2$ must use \(t \gtrsim n\) matvecs.
\end{theorem}
To the best of our knowledge, \cref{thm:main-deterministic} demonstrates the first provable advantage of randomized algorithms over deterministic ones for solving generic linear systems in a natural computational model.
We prove the result in \cref{sec:intro-det-lower-bound} by showing that deterministic methods cannot do better than a Krylov method run on the normal equations, $\mA^\top\mA\vx = \mA^\top\vb$.
We then combine this with a polynomial inapproximability result showing that at least a logarithmic dependence on the condition number of $\mA$ is unavoidable.
This condition number dependence is quite unexpected when dealing with backward error, since by definition one should be able to perturb away from an ill-conditioned problem.
Yet, our result implies that finding such a perturbation is information-theoretically impossible without randomness.

\subsection{
Contribution 3: Improved complexity of symmetric systems
}
While \cref{thm:main-deterministic} rules out condition-number-independent deterministic algorithms for solving general linear systems to small backward error, \cite{derezinski2026towards} at least shows that a mild condition number dependence \emph{is} achievable. Indeed, they prove that for every $n\times n$ linear system $\mA\vx=\vb$ with condition number $\kappa$, an $\eps$ backward error solution $\widetilde \vx$ can be recovered from $\cO(\log (\kappa)/\eps)$ matvecs with $\mA$ and $\mA^\top$: specifically, from the Krylov subspace of degree $\cO(\log(\kappa)/\eps)$ for the normal equations $\mA^\top\mA\vx = \mA^\top\vb$. This is a significant improvement over the polynomial dependence on $\kappa$ required to obtain small forward error \cite{derezinski2026matrix}. 

Interestingly, we show that, if we restrict $\mA$ to be symmetric, then the condition number dependence can actually be removed entirely. Specifically, we are able to attain $\cO(1/\eps)$ deterministic matvec complexity:
\begin{theorem}[Solving symmetric systems to backward error deterministically]
    \label{thm:main-minberr-symmetric}
    Let \(\mA\in\bbR^{n \times n}\) be a symmetric matrix.
    We can solve the linear system \(\mA\vx=\vb\) to backward error \(\eps\) using \(t\le 5/\eps\) deterministic matvecs with \mA. Moreover, this is optimal: Any (potentially randomized) algorithm requires \(t\geq (e^{-1}-o(1))/\eps\) matvecs to solve a symmetric linear system to backward error $\eps$.
\end{theorem}
\cref{thm:main-minberr-symmetric} improves on the randomized $\cO(\log(n)/\eps)$ matvec bound for general linear systems from \cite{derezinski2026towards}. Moreover, the algorithm can be seen to run in $\cO(n^2/\eps)$ deterministic time, which for small constant $\eps$, is faster than deterministic system solvers based on fast matrix multiplication \cite{demmel2007fast,dupont2026improving}. 

\Cref{thm:main-minberr-symmetric} is proven in \Cref{sec:minberr-symmetric-upper} (upper bound) and \Cref{sec:symmetric-lower-bounds} (lower bound). Like \cite{derezinski2026towards}, our approach is based on minimizing backward error over a Krylov subspace. As is typical for Krylov methods, the analysis thus reduces to a polynomial approximation problem \cite{greenbaum1997iterative}. However, unlike existing Krylov analyses,
we show that, while no fixed polynomial exists that attains our required approximation bound, it is always possible to choose a polynomial that works for any given right-hand side vector $\vb$.

\section{Preliminaries}

We write scalars in italic, vectors in bold lowercase, and matrices in bold uppercase.
Universal constants may also be written as upright non-italics.
The sets of real numbers, real vectors, and real matrices are denoted \(\bbR\), \(\bbR^n\), and \(\bbR^{n\times d}\), respectively.
The set of all unit vectors in \(\bbR^n\) is denoted \(\bbS^{n-1}\).
Other sets are written in calligraphic letters.
We use the standard big-Oh symbols \(\cO\), \(\Omega\), \(\Theta\) and \(o\).
We write \(f \lesssim g\) if there exists a universal constant \(\rC>0\) such that \(f \leq \rC g\), and define \(f \gtrsim g\) analogously.
We write \(f \asymp g\) if \(f \lesssim g\) and \(f \gtrsim g\).

We write probability as \(\bbP[\cdot]\), expectation as \(\E[\cdot]\), equality in distribution as \(\overset{\mathrm{dist}}=\), and the indicator of an event \cE as \(\mathbbm1_{\cE}\).
The (multivariate) Gaussian distribution is denoted \(\cN(\mu,\sigma)\).
We say that a matrix is a Gaussian matrix if it has iid \(\cN(0,1)\) entries.
The chi distribution with parameter \(k\) is \(\chi_k\), the uniform distribution on interval \([a,b]\) is \(\textsc{Unif}[a,b]\), and the point mass distribution at \(x\) is \(\delta_x\).
For a set \(\cS\subseteq\bbR\), the supremum norm is \(\norm{f}_{\cS} = \sup_{x\in\cS}|f(x)|\) and the \(L_2\) norm on \cS is \(\norm{f}_{L_2(\cS)} = (\int_\cS |f(x)|^2 dx)^{1/2}\).
For a measure \(\mu\), the weighted \(L_2\) norm is \(\norm{f}_{L_2(\mu)} = (\int |f(x)|^2 d\mu(x))^{1/2}\).
The distance from a point to a set is \(\operatorname{dist}(\vx,\cS) \defeq \min_{\vy\in\cS}\norm{\vx-\vy}_2\).

The standard basis vectors in \(\bbR^n\) are \(\ve_1,\ldots,\ve_n\), and the identity matrix is \mI or \(\mI_n\). 
The \(L_2\) norm of a vector is \(\norm{\vv}_2\), and the standard inner product is \(\langle \vu, \vv \rangle = \vu^\top\vv\).
The singular values of a matrix \(\mA\in\bbR^{n \times d}\) are sorted in decreasing order: \(\sigma_{\max}(\mA) = \sigma_1(\mA) \geq \ldots \geq \sigma_{\min\{n,d\}}(\mA) = \sigma_{\min}(\mA)\).
The condition number of \mA is \(\cond(\mA) \defeq \sigma_{\max}(\mA) / \sigma_{\min}(\mA)\).
The operator norm is \(\norm{\mA}_2\), the Frobenius norm is \(\norm\mA_{\rm F}\), and the trace is \(\tr(\mA)\).
The eigenvalues of a symmetric matrix \(\mA\in\bbR^{n \times n}\) are sorted in decreasing order: \(\lambda_{\max}(\mA) = \lambda_1(\mA) \geq \ldots \geq \lambda_n(\mA) = \lambda_{\min}(\mA)\).
A matrix is positive semidefinite (psd) if it is symmetric and has non-negative eigenvalues.
We denote this as \(\mA\succeq\mat0\).

For a linear system \(\mA\vx=\vb\), a vector \(\widetilde\vx\) has forward error \(\norm{\widetilde\vx - \mA^{-1}\vb}_2 / \norm{\mA^{-1}\vb}_2\), relative residual \(\norm{\mA\widetilde\vx-\vb}_2 / \norm{\vb}_2\), and backward error
\begin{equation*}
    \mathrm{berr}(\mA,\vb,\widetilde\vx)
    \defeq \min\left\{
        \eps \geq 0
        ~:~
        (\mA+\mDelta)\widetilde\vx=\vb
        \text{ for some \(\mDelta\) with } \norm{\mDelta}_2\leq\eps\norm\mA_2
    \right\}
    = \frac{\norm{\mA\widetilde\vx-\vb}_2}{\norm\mA_2\norm{\widetilde\vx}_2},
\end{equation*}
where the latter equality is a folklore fact in numerical analysis.
The Krylov subspace of degree \(t\) for matrix \mA and vector \vb is
\[
    \cK_t(\mA,\vb) \defeq \Span\bigl\{\vb,~\mA\vb,~\mA^2\vb,~\cdots,~\mA^t\vb\bigr\}.
\]

\section{Technical overview}

Here, we give a broad technical overview of our results and the techniques that underlie them. Our analysis develops and uses a range of tools from random matrix theory, information-based complexity, and polynomial approximation, which together form a basis for studying backward error and perturbation-based algorithms. We review these tools as they come up in discussing each of the individual results.

\subsection{Smoothed analysis lower bound}

In this section, we overview the proof of \cref{thm:main-smoothed}, i.e., that in the smoothed analysis model, solving a linear system to constant forward error with constant probability requires \(\Omega(n)\) matvecs. Remarkably, we show that this is true regardless of the distribution of the perturbation, as long as it is oblivious to the input data and scaled to be smaller than the input matrix. To show this, we consider the linear system \(\mA\vx=\vb\) where \(\mA\in\bbR^{n \times n}\) has iid standard Gaussian entries and \vb is arbitrary. \begin{theorem}[Smoothed lower bound]\label{thm:smoothed-lower-bound}
Let \(\mA\in\bbR^{n\times n}\)
have iid \(\cN(0,1)\) entries, let \(\mDelta\in\bbR^{n\times n}\)
be independent of \(\mA\) with
\(\norm{\mDelta}_2\le\sqrt n/3\) almost surely, and set
\(\widehat\mA\defeq\mA+\mDelta\).
There is a universal constant $p\in(0,1)$ such that any algorithm with matvec access to \(\widehat\mA\) and \(\widehat\mA^\top\) requires \(\Omega(n)\) matvecs to output a vector \(\widetilde\vx\) where
    \[
        \norm{\widetilde\vx - \widehat\mA^{-1}\vb}_2 \leq \frac1{20} \norm{\widehat\mA^{-1}\vb}_2
        \quad\text{with prob.~} \geq p.
    \]
\end{theorem}
Note that, for a Gaussian $\mA$, with high probability $\norm{\mA}_2 \approx 2\sqrt n$, so the size of perturbation $\mDelta$ is within a constant backward error of the size of the input, i.e., about as large as one might consider without the noise drowning out the signal. Our proof also shows that the knowledge of $\mDelta$ does not affect the lower bound, so the result still applies if the perturbation is explicitly generated and applied by the algorithm. To recover \cref{thm:main-smoothed} from \cref{thm:smoothed-lower-bound}, we scale the input by $\Theta(1/\sqrt n)$ so that its norm is $\leq 1$ with high probability.

\paragraph{Condition number of perturbed Gaussian.} The starting point behind our lower bound construction is the observation that an $n\times n$ Gaussian matrix $\mA$ has condition number $\Theta(n)$ with high probability, which means that we would expect a standard matvec algorithm (e.g., a Krylov method like LSQR) to take $\Omega(n)$ matvecs to solve it. Naturally, this is not sufficient to show a lower bound over all matvec algorithms, and in fact, to our knowledge \cref{thm:smoothed-lower-bound} is the first such hardness result for a Gaussian linear system even if one omits the perturbation. The next step in the reasoning is to address whether some oblivious perturbation can improve the condition number of a Gaussian matrix. The following result shows that it cannot, by controlling the smallest singular value of $\widehat \mA$.

\begin{theorem}[The minimum singular value of a perturbed Gaussian matrix]\label{thm:gaussian-near-singularity}
Let $n\ge 8$, \(\mA\in\bbR^{n\times n}\)
have iid \(\cN(0,1)\) entries, let \(\mDelta\in\bbR^{n\times n}\)
be independent of \(\mA\) with
\(\norm{\mDelta}_2\le\sqrt n/2\) almost surely, and set
\(\widehat\mA\defeq\mA+\mDelta\). Then, for all \(s \geq 1\),
    \[
        \sigma_{\min}(\widehat\mA) \leq \frac1{s \sqrt n}
        \quad
        \text{with prob. } \gtrsim \frac{1}{s^2}.
    \]
\end{theorem}
Since the largest singular value of $\widehat \mA$ is not significantly affected by the perturbation, \cref{thm:gaussian-near-singularity} shows that the condition number of $\widehat \mA$ is $\Theta(n)$ with constant probability. With some effort, this result can be derived by building on recent spectral density estimates for shifted Ginibre ensembles, e.g., based on the matrix resolvent concentration inequalities of Cipolloni, Erd\H{o}s, and Schr\"oder \cite{cipolloni2022density}. Instead, in \cref{sec:gaussian-near-singularity} we give a much more elementary proof that does not require any recent results from random matrix theory, and should be of independent interest. Curiously, the high-level structure of this argument mirrors the smoothed analysis work of Sankar, Spielman, and Teng \cite{sankar2006smoothed}, who show a matching lower bound on the smallest singular value of a shifted Gaussian matrix.

\paragraph{Hidden Gaussian theorem.} Clearly, a condition number bound is not sufficient to prove hardness against all matvec algorithms. To address this, we follow the framework of a \emph{hidden matrix theorem} \cite{simchowitz2018randomized,braverman2020gradient,amsel2026fixed,derezinski2026matrix}, where after the algorithm observes the outcome of its queries, we show that there is still a random submatrix of a transformed input that is entirely unobserved by the algorithm. Here, unlike prior work, we must consider an \emph{asymmetric} matrix \mA and are allowed to compute matvecs with both \(\mA^\top\) and \mA. In particular, in \cref{thm:hidden-gaussian} we show that after observing $t$ matvecs of each kind, the input matrix can be decomposed as:
\[
        \mR^\top\mA\mS = \bmat{\mB_{11} & \mB_{12} \\ \mB_{21} & \mG},
\]
where $\mR$ and $\mS$ form a change of basis and $\mG\in\bbR^{(n-t)\times (n-t)}$ is a Gaussian matrix independent of the past.  We then invoke \cref{thm:gaussian-near-singularity} to show that this submatrix is still sufficiently ill-conditioned. Finally, we complete the argument via a reduction from the task of approximating the inverse Frobenius norm $\norm{\widehat\mA^{-1}}_{\rm F}^2$ to solving linear systems with $\widehat\mA$.

\subsection{Deterministic lower bound}
\label{sec:intro-det-lower-bound}

In this section, we explain the proof of \cref{thm:main-deterministic}, our matvec lower bound against deterministic algorithms that solve a general linear system to small backward error.
The specific lower bound we show is that any deterministic solver must have matvec complexity depending on the condition number of \mA.
\begin{theorem}[Deterministic backward error solvers depend on the condition number]
    \label{thm:det-cond-lower-bound}
Every deterministic matvec algorithm that can solve any linear system \(\mA\vx=\vb\) of condition number at most \(\cond(\mA)\leq\kappa\) to backward error \(\eps\) must use \(\gtrsim\min\{\log(\kappa)/\eps,~n\}\) matvecs in the worst case.
\end{theorem}
Our lower bound matches the upper bound of \cite{derezinski2026towards} up to constant factors, showing that the deterministic matvec complexity of solving a linear system with condition number $\kappa$ to $\eps$ backward error is $\Theta(\log(\kappa)/\eps)$. On the other hand, if we consider  $n\times n$ linear systems without any assumptions on the condition number, then this proves a \(\Theta(n)\) deterministic matvec complexity. 

The proof works in three major steps.
First, we show that any deterministic method can be reduced to a Krylov method that runs on the normal equations.
To define this, for a matrix \(\mA\), let \(t_{\rm ne}(\mA,\vb)\) be the (odd) degree \(t=2k+1\) needed for a vector in the Krylov subspace \(\cK_{k}(\mA^\top\mA,\mA^\top\vb)\) for the normal equations to have small backward error:
\begin{equation}
    \label{eq:t-ne}
    t_{\rm ne}(\mA,\vb) \defeq \min\left\{
        2k+1
        ~:~
        k \geq 0,\quad \exists \vx\in\cK_{k}(\mA^\top\mA,\mA^\top\vb) \text{ such that } \mathrm{berr}(\mA,\vb,\vx) \leq \eps
    \right\}.
\end{equation}
That is, a method which returns a vector from the Krylov subspace associated with the normal equations must compute at least \(t_{\rm ne}(\mA,\vb)\) matvecs.
We show the following reduction.
\begin{theorem}[Deterministic methods cannot do better than Krylov on the normal equations]
    \label{thm:det-lower-bound-to-krylov}
    Let \(\cF \subseteq \bbR^{n \times n}\) be a set of invertible matrices closed under orthogonal actions: \(\mU^\top\mA\mV\in\cF\) for all orthogonal \(\mU,\mV\in\bbR^{n \times n}\).
    Fix any \(\vb \in \bbR^n\).
    Let \(\Phi(\mA,\vb)\) be any deterministic matvec algorithm that solves linear systems to backward error \(\eps\).
    Let \(t_{\Phi}(\mA,\vb)\) denote the number of matvecs that \(\Phi\) uses to solve \(\mA\vx=\vb\) to backward error \(\eps\).
    Then,
    \[
        \sup_{\mA\in\cF} t_{\Phi}(\mA,\vb) \geq \frac14 \left( \sup_{\mA\in\cF} t_{\rm ne}(\mA,\vb) - 1 \right).
    \]
\end{theorem}
The proof of \cref{thm:det-lower-bound-to-krylov}, given in \cref{sec:det-lower-bounds},
is a generalization of resisting oracle arguments from the \emph{information-based complexity theory} literature \cite{chou1987optimality,traub2003information}.
The existing work proved lower bounds for deterministic methods on different classes of \emph{symmetric} matrices.
Our proof is modeled after a result by Chou \cite{chou1987optimality}, though it is significantly generalized to handle non-symmetric matrices and a range of error metrics.
For our purposes this is a strength, as showing that deterministic methods are provably weaker than randomized methods requires an argument that treats deterministic methods differently than randomized~ones.

With \cref{thm:det-lower-bound-to-krylov} in place, we need to lower bound the worst-case complexity of a Krylov method on a general matrix \mA.
We can relate this to a question about polynomial approximation in a weighted \(L_2\) norm.
\begin{lemma}[From Krylov on the normal equations to odd polynomials]
    Let \(\mA\in\bbR^{n \times n}\) and \(\vb\in\bbR^n\).
    Expand the SVD \(\mA=\mU\mSigma\mV^\top\) with singular values \(\sigma_1 \geq \ldots \geq \sigma_n\).
    Fix any odd number \(t \geq 1\) and expand \(t=2k+1\).
    Then, the minimum backward error achievable by any vector in the Krylov subspace for the normal equations is
    \label{lem:normal-to-odd-poly}
    \begin{equation}
        \label{eq:normal-to-odd-poly}
        \min_{\vx\in\cK_k(\mA^\top\mA,\mA^\top\vb)}
        (\mathrm{berr}(\mA,\vb,\vx))^2
        = \min_{\substack{\deg(p) \leq t:\\p\text{ is odd}}}
            \frac{\sum_{i=1}^n w_i (\sigma_i p(\sigma_i) - 1)^2}
                 {\sigma_1^2 \sum_{i=1}^n w_i p(\sigma_i)^2},
    \end{equation}
    where \(w_i = [\mU^\top\vb]_i^2\).
\end{lemma}
\begin{proof}
    A vector \(\vx\in\cK_k(\mA^\top\mA,\mA^\top\vb)\) if and only if \(\vx=\mV p(\mSigma)\mU^\top\vb\) for an odd polynomial of degree \(t\).
    By the orthogonal invariance of the \(L_2\) norm, we have for every \(\vx\in\cK_k(\mA^\top\mA,\mA^\top\vb)\) that
    \[
        (\mathrm{berr}(\mA,\vb,\vx))^2
        = \frac{\norm{\mA\vx-\vb}_2^2}{\norm{\mA}_2^2 \norm{\vx}_2^2}
= \frac{\norm{(\mSigma p(\mSigma)-\mI)\mU^\top\vb}_2^2}{\norm{\mSigma}_2^2 \norm{p(\mSigma)\mU^\top\vb}_2^2}
        = \frac{\sum_{i=1}^n w_i(\sigma_i p(\sigma_i)-1)^2}{\sigma_1^2 \sum_{i=1}^n w_i p(\sigma_i)^2}.
    \]
    We conclude by minimizing over \(\vx\in\cK_k(\mA^\top\mA,\mA^\top\vb)\), and thereby over odd polynomials \(p\) of degree \(t\).
\end{proof}

\begin{figure}
    \centering
    \resizebox{0.8\linewidth}{!}{\begin{tikzpicture}[x=1cm,y=1cm]
\path[use as bounding box] (0,3.40) rectangle (16.2,8.35);
\fill[black!6] (8.20,3.58) rectangle (14.88,7.80);
        \draw[guide] (8.20,4.15)--(8.20,7.80);
        \draw[guide] (14.88,4.15)--(14.88,7.80);
        \draw[axis] (1.18,4.15)--(15.62,4.15) node[right] {$x$};
        \draw[axis] (1.45,3.53)--(1.45,7.96) node[above] {$p(x)$};
        \draw[guide] (1.45,7.02)--(3.25,7.02);
        \draw[guide] (3.25,4.15)--(3.25,7.02);
        \draw[curve] (1.45,4.15)
         .. controls (1.75,4.15) and (2.60,7.02) .. (3.25,7.02)
         .. controls (4.00,7.02) and (4.60,5.12) .. (5.70,4.75)
         .. controls (6.75,4.39) and (7.22,4.40) .. (8.20,4.32)
         .. controls (8.70,4.29) and (9.00,3.98) .. (9.63,4.03)
         .. controls (10.30,4.09) and (10.80,4.41) .. (11.45,4.36)
         .. controls (12.15,4.31) and (12.55,3.91) .. (13.18,3.98)
         .. controls (13.88,4.06) and (14.40,4.38) .. (14.88,4.27);
        \fill (3.25,7.02) circle (1.5pt);
        \fill (1.45,4.15) circle (1.5pt);
        \node[anchor=west] at (3.50,7.50) {$p(1/\kappa)\gtrsim1/\eps$};
        \draw[edge] (4.02,7.20)--(3.43,7.06);
        \node[align=center] at (11.55,6.70) {Small on average\\over $[2\eps,1]$};
\node[minor,rotate=66] at (2.00,5.80) {rapid growth};
        \draw (3.25,4.15)--(3.25,4.02);
        \draw (8.20,4.15)--(8.20,4.02);
        \draw (14.88,4.15)--(14.88,4.02);
        \node[below] at (3.25,3.95) {$1/\kappa$};
        \node[below] at (8.20,3.95) {$2\eps$};
        \node[below] at (14.88,3.95) {$1$};
\node[anchor=east] at (1.35,4.38) {$p(0)=0$};

\end{tikzpicture}}
    \caption{
        Illustration of hardness of approximation in backward error with odd polynomials (\cref{thm:odd-poly-inapprox}).
         The requirement that an odd polynomial be simultaneously small on $[2\eps,1]$ and large at $1/\kappa$ forces a degree of $\Omega(\log(\kappa)/\eps)$
}
    \label{fig:odd-polynomial}
\end{figure}
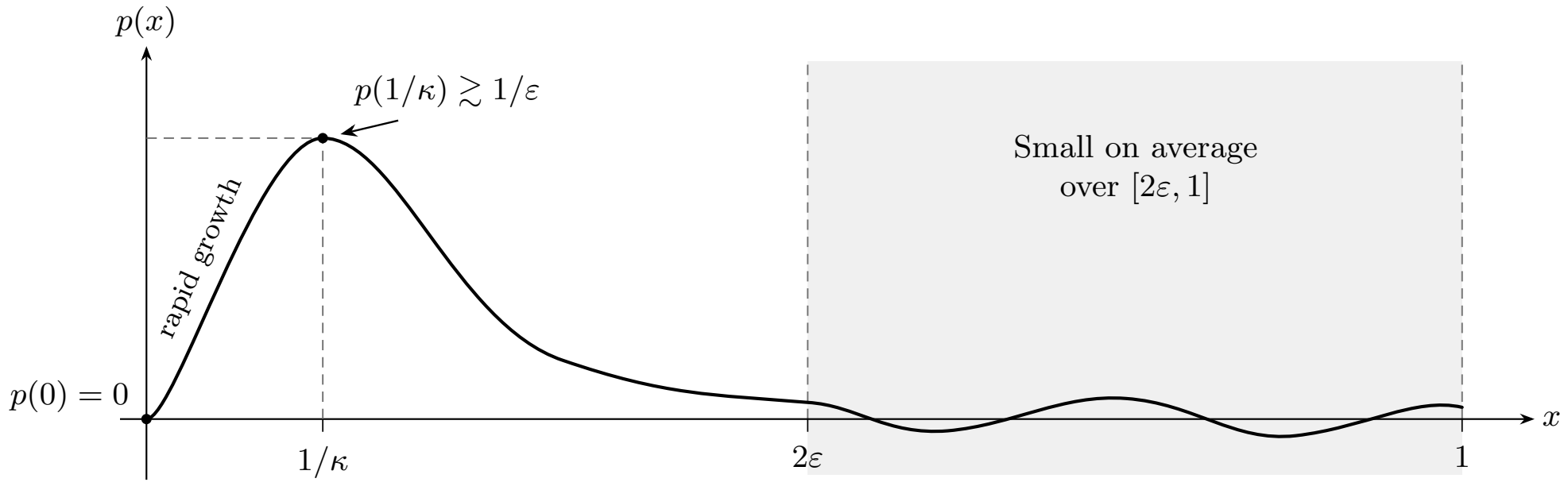

In particular, if there exists a single spectrum \(\sigma_1,\ldots,\sigma_n\in[1/\kappa,1]\) and weights \(w_i\geq0\) such that no odd polynomial can make the right-hand side of \cref{eq:normal-to-odd-poly} small, then \cref{lem:normal-to-odd-poly} shows no vector in the Krylov subspace achieves small error on some corresponding matrix \mA and vector \vb.
From there, \cref{thm:det-lower-bound-to-krylov} shows that no deterministic method may achieve small backward error.
Hence, the only remaining step is to prove such a polynomial inapproximability claim.
\begin{theorem}[Hardness of approximation with odd polynomials]
    \label{thm:odd-poly-inapprox}
    Fix \(n \gtrsim 1\), \(\kappa \gtrsim 1\), and \(\kappa^{-0.1}\lesssim \eps \lesssim 1\).
    Then, there exist weights \(w_1,\ldots,w_n\geq0\) and nodes \(\sigma_1,\ldots,\sigma_n\in[1/\kappa,1]\) such that any odd polynomial \(p\) with
    \begin{equation}
        \label{eq:odd-approx-goal}
        \frac{\sum_{i=1}^n w_i (\sigma_i p(\sigma_i) - 1)^2}
        {\sigma_1^2 \sum_{i=1}^n w_i p(\sigma_i)^2}
        \leq\eps^2
    \end{equation}
    must have degree \(t \gtrsim \min\{n,\log(\kappa)/\eps\}\).
\end{theorem}
We prove \cref{thm:odd-poly-inapprox} in \cref{sec:odd-poly-inapprox}.
The argument is a consequence of several basic tools in polynomial approximation theory: the Bernstein inequality, the pointwise Remez inequality, and the Nikolskii inequality.
Conceptually, the argument places the weights \(w_i\) in a way to ensure that any \(p\) satisfying \cref{eq:odd-approx-goal} must satisfy \(p(1/\kappa) \gtrsim 1/\eps\) and be small on average on the interval \([2\eps,1]\), as shown in \cref{fig:odd-polynomial}.
Further, any odd polynomial has \(p(0)=0\).
By chaining together the Bernstein, pointwise Remez, and Nikolskii inequalities, we show that any odd polynomial satisfying these two constraints must have degree \(t \gtrsim \min\{n, \log(\kappa)/\eps\}\).

Notably, if we were not forced to take \(p\) to be odd (a consequence of solving the normal equations), then no such lower bound would hold.
Indeed, the opposite is true: the analysis of algorithms minimizing backward error on psd systems in \cite{derezinski2026towards} using degree \(t \lesssim 1/\sqrt\eps\) implies that 
there exists a polynomial \(q\) which does satisfy \cref{eq:odd-approx-goal} with degree independent of the condition number.

\subsection{Deterministic upper bound for symmetric systems}

\cref{thm:main-minberr-symmetric} argues that if \mA is symmetric, then we can use \(\cO(1/\eps)\) deterministic matvecs with \mA to solve \(\mA\vx=\vb\) to \(\eps\) backward error. What enables us to show this is that, due to symmetry, we can avoid relying on the normal equations, and can instead consider the Krylov subspace of the system itself.
Our algorithmic approach follows that of \cite{derezinski2026towards} for solving psd linear systems: perform \(t\) matvecs to build a Krylov subspace \(\cK_t(\mA,\vb)\) and return the vector \(\widetilde\vx_{\rm minberr}\in\cK_t(\mA,\vb)\) that minimizes the backward error.

For psd matrices, \cite{derezinski2026towards} shows that \(t \lesssim 1/\sqrt\eps\) matvecs suffice, but for symmetric indefinite matrices their analysis fundamentally breaks down. To see why, let us examine their reduction from backward error to a polynomial approximation guarantee in the $L_\infty$ norm:
\begin{equation*}
    \min_{\vx\in\cK_t(\mA,\vb)} \mathrm{berr}(\mA,\vb,\vx)
    \leq \min_{\deg(p) \leq t} \max_{x\in[0,1]} \left|x - \frac1{p(x)}\right|
    \qquad\text{for psd \mA}.
\end{equation*}
The right-hand side features an out-of-the-box polynomial approximation problem that has a long history in approximation theory \cite{levin1988degree}. If we extend this to indefinite matrices, the same argument yields:
\begin{equation}
    \label{eq:indef-minberr-reduction}
\min_{\vx\in\cK_t(\mA,\vb)} \mathrm{berr}(\mA,\vb,\vx)
    \leq \min_{\deg(p) \leq t} \max_{x\in[-1,1]} \left|x - \frac1{p(x)}\right|
    \qquad\text{for symmetric \mA},
\end{equation}
where we now have \(x\in[-1,1]\) instead of \(x\in[0,1]\).
Unfortunately, this optimization problem is now impossible to bound effectively because of the behavior for \(x\) near zero.
Fix some small parameter \(\delta>0\).
For \(x=\delta\), we must have that \(1/p(\delta)\) is small and positive, so \(p(\delta)\) must be large and positive.
For \(x=-\delta\), we conversely need that \(p(-\delta)\) is large and negative.
These two forces contradict each other, meaning that no polynomial makes the right-hand side of \cref{eq:indef-minberr-reduction} small.

\begin{figure}
    \centering
    \resizebox{0.8\linewidth}{!}{\begin{tikzpicture}[x=1cm,y=1cm]
\path[use as bounding box] (0,2.35) rectangle (16.2,5.48);
        \draw[panel] (0,2.35) rectangle (7.80,5.48);
        \node[font=\polyfont\bfseries] at (3.90,5.12) {(a) Little mass near zero};
        \fill[black!6] (3.20,2.60) rectangle (4.60,4.48);
        \draw[axis] (.65,3.54)--(7.15,3.54);
        \draw[axis] (3.90,2.60)--(3.90,4.55);
        \draw[black!50,densely dashed,line width=.6pt,domain=-1:-.245,samples=60]
         plot ({3.90+2.84*\x},{3.54+.22/\x});
        \draw[black!50,densely dashed,line width=.6pt,domain=.245:1,samples=60]
         plot ({3.90+2.84*\x},{3.54+.22/\x});
\draw[curve] (1.06,3.32)
         .. controls (1.65,3.29) and (2.48,3.13) .. (2.95,2.91)
         .. controls (3.33,2.72) and (3.63,2.97) .. (3.90,3.54)
         .. controls (4.17,4.11) and (4.47,4.36) .. (4.85,4.17)
         .. controls (5.32,3.95) and (6.15,3.79) .. (6.74,3.76);

        \draw[panel] (8.40,2.35) rectangle (16.20,5.48);
        \node[font=\polyfont\bfseries] at (12.30,5.12) {(b) Substantial mass near zero};
        \fill[black!6] (11.60,2.92) rectangle (13.00,4.48);
        \draw[axis] (9.05,2.96)--(15.55,2.96);
        \draw[axis] (12.30,2.92)--(12.30,4.55);
        \draw[curve] (9.46,3.00)
         .. controls (9.81,3.09) and (10.04,3.10) .. (10.32,2.99)
         .. controls (10.68,2.97) and (10.91,3.23) .. (11.22,3.43)
         .. controls (11.72,3.83) and (11.86,4.37) .. (12.30,4.37)
         .. controls (12.74,4.37) and (12.88,3.83) .. (13.38,3.43)
         .. controls (13.69,3.23) and (13.92,2.97) .. (14.28,2.99)
         .. controls (14.56,3.10) and (14.79,3.09) .. (15.14,3.00);
        \end{tikzpicture}}
    \caption{Illustration of the polynomials used in the symmetric upper bound (\cref{thm:indefinite-poly-approx}). There are two cases: (a) When $\mu$ places little mass near zero, then we use a polynomial that approximates $1/x$ well far from the origin; (b) When $\mu$ places much of its mass near zero, then we use a simple non-negative polynomial that is large near zero.}
    \label{fig:symmetric}
\end{figure}
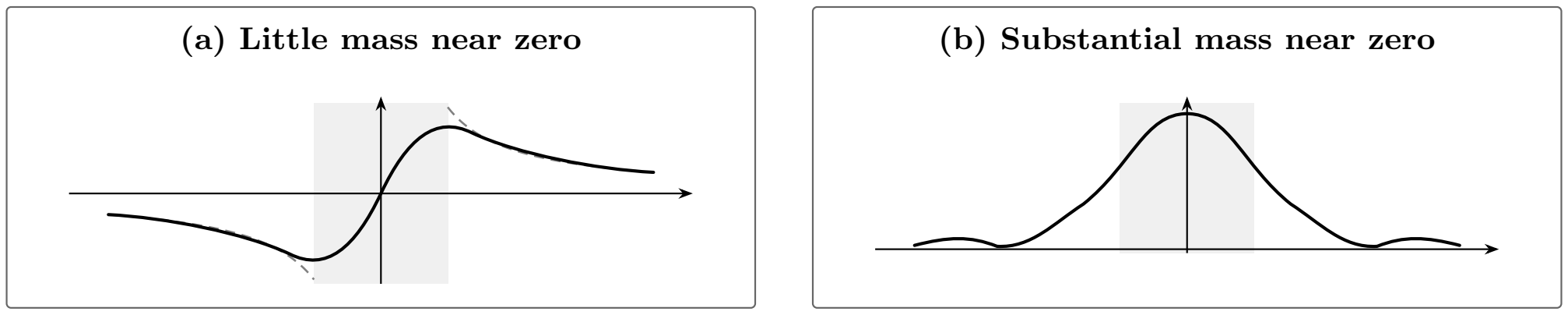

So, to prove the result requires a more careful analysis.
Without loss of generality, scale \mA to have unit operator norm.
Using the same argument as in \cref{lem:normal-to-odd-poly}, we can write
\begin{equation}
    \label{eq:sym-minber-reduction}
    \min_{\vx\in\cK_t(\mA,\vb)} (\mathrm{berr}(\mA,\vb,\vx))^2
    = \min_{\deg(p) \leq t} \frac{\sum_{i=1}^n w_i (\lambda_i p(\lambda_i)-1)^2}{\sum_{i=1}^n w_i p(\lambda_i)^2},
\end{equation}
where \(\lambda_1,\ldots,\lambda_n\) are the eigenvalues of \mA and \(w_i\geq0\) depend on the eigenvectors of \mA and the vector \vb.
This polynomial approximation problem is fundamentally comparing two vectors in a weighted \(L_2\) norm, in contrast to the right-hand side of \cref{eq:indef-minberr-reduction} which is measured in the \(L_\infty\) norm.
It turns out that going from the \(L_2\) to \(L_\infty\) norm is suboptimal 
when analyzing symmetric (but not psd) problems.
\begin{theorem}[Polynomial approximation in the \(L_2\) norm]
    \label{thm:indefinite-poly-approx}
    Fix \(\eps>0\).
    Let \(\mu\) be a probability distribution supported on \([-1,1]\), and let \(\norm{f}_{L_2(\mu)} \defeq (\int |f(x)|^2 d\mu(x))^{1/2}\) be the \(\mu\)-weighted \(L_2\) norm.
    There exists a polynomial \(p\) of degree at most \(k=\lceil 5/\eps \rceil\) such that
    \begin{equation}
        \label{eq:symmetric-mu-guarantee}
        \norm{xp-1}_{L_2(\mu)} \leq \eps \norm{p}_{L_2(\mu)}.
    \end{equation}
\end{theorem}
We prove \cref{thm:indefinite-poly-approx} in \cref{sec:minberr-symmetric-upper}.
The combination of \cref{eq:sym-minber-reduction} and \cref{thm:indefinite-poly-approx} immediately proves the guarantee for symmetric matrices (first part of \cref{thm:main-minberr-symmetric}).
We prove \cref{thm:indefinite-poly-approx} by case analysis over the distribution \(\mu\).
If \(\mu\) places only a small amount of mass near zero, then we can use a relatively off-the-shelf \(L_\infty\) norm polynomial approximation to the function \(x \mapsto 1/x\) to prove the claim.
This should not be surprising: if \(\mu\) places little mass near zero then we can avoid thinking about the region where approximating the function \(1/x\) grows large; see \cref{fig:symmetric}(a).

The perhaps more surprising result is that if \(\mu\) concentrates much of its mass near zero, then we can achieve \cref{eq:symmetric-mu-guarantee} by taking \(p\) to be a nonnegative bump function.
This polynomial is by no means a conventional approximation to the function \(x\mapsto1/x\).
To understand this better, consider \cref{eq:symmetric-mu-guarantee} normalized by \(\norm{p}_{L_2(\mu)}=1\).
The left-hand side depends on the term \(x p(x)\), which is small so long as either \(x\) is small or \(p(x)\) is small.
So, if \(\mu\) places much of its mass near zero, then it suffices to also make \(p(x)\) place all of its mass near zero; see \cref{fig:symmetric}(b).

\subsection{Lower bounds for symmetric and psd systems}

In this section, we outline the proof of the lower bound side of \cref{thm:main-minberr-symmetric}, i.e., the matvec lower bound against randomized algorithms solving symmetric systems. In fact, with no additional effort, we also achieve a matching lower bound for the complexity of solving psd systems to small backward error.
To motivate our lower bound technique, note that the matvec complexities in both the symmetric and psd cases follow a pattern.
Here, \(\kappa\) denotes the condition number of the linear system, and recall that a vector \(\widetilde\vx\) has \(\eta\) relative residual if \(\norm{\mA\widetilde\vx-\vb}_2 \leq \eta \norm{\vb}_2\).
\begin{itemize}
    \item For symmetric matrices, \(\Theta(\kappa\log(1/\eta))\) matvecs are needed to achieve \(\eta\) relative residual, and \(\Theta(1/\eps)\) are needed to achieve \(\eps\) backward error.
    \item For psd matrices, \(\Theta(\sqrt\kappa\log(1/\eta))\) matvecs are needed to achieve \(\eta\) relative residual, and \(\Theta(1/\sqrt\eps)\) are needed to achieve \(\eps\) backward error.
\end{itemize}
In both cases, the complexity of achieving backward error \(\eps\) equals the complexity of achieving relative residual \(\eta=0.5\) if we set \(\kappa = 1/\eps\).
This is not a coincidence, but instead a systematic relationship.
A folklore fact in numerical analysis says that, up to a small constant, the relative residual is bounded by the product of the condition number and the backward error.
In order to sharpen our constants, we prove a slightly stronger version of this fact:
\begin{lemma}[Backward-to-residual reduction]
    \label{lem:bwd-to-res-reduction}
    Let \(\eps > 0\) and let \mA have condition number at most \(\kappa\).
    Fix any vectors \vb and \(\widetilde\vx\) such that \(\mathrm{berr}(\mA,\vb,\widetilde\vx) \leq \eps\).
    Then, using a single matvec with \mA, we can construct a vector \(\widehat\vx\) such that \(\norm{\mA\widehat\vx-\vb}_2\leq\kappa\eps\norm{\vb}_2\).
\end{lemma}
This lemma, proven in \cref{sec:symmetric-lower-bounds}, shows that any method which achieves \(\eps\) backward error can be used to produce a constant residual iterate on systems of condition number \(\kappa<1/\eps\).
So, since any algorithm which solves a psd system of condition number \(\kappa\) to constant residual error must use \(\Omega(\sqrt\kappa)\) matvecs \cite{derezinski2026matrix}, we conclude that solving a psd system to \(\eps\) backward error needs \(\Omega(1/\sqrt\eps)\) matvecs.
The analogous argument covers symmetric systems.
The formal version of this argument, proving the lower bound of \cref{thm:main-minberr-symmetric}, is also given in \cref{sec:symmetric-lower-bounds}.
 
\section{Hardness of smoothed analysis}
\label{sec:smoothed-lower-bound}

Here, we complete the proof of \cref{thm:smoothed-lower-bound}. To simplify the argument, we will strengthen our algorithm from a basic matvec algorithm to a \emph{paired matvec algorithm}, which at every timestep computes a matvec with both \mA and \(\mA^\top\):
\[
    (\vu_t,\vv_t) \to \textsc{oracle} \to (\mA^\top\vu_t, \mA\vv_t).
\]
We will call \(\vu_t\) the \textit{left queries} and \(\vv_t\) the \textit{right queries}.
Our goal is to show that any paired matvec algorithm must compute \(\Omega(n)\) queries.
To start, we prove a hidden matrix theorem, a result analogous to many in the literature \cite{simchowitz2018randomized,braverman2020gradient,amsel2026fixed,derezinski2026matrix}.
However, unlike existing results, here we focus on an asymmetric matrix \mA.

\begin{theorem}[Hidden Gaussian theorem]
    \label{thm:hidden-gaussian}
    Let \(\mA\in\bbR^{n \times n}\) have iid \(\cN(0,1)\) entries.
Suppose a paired matvec algorithm computes (possibly adaptive) left queries \(\vu_1,\ldots,\vu_t\) and right queries \(\vv_1,\ldots,\vv_t\), resulting in a transcript of matvecs \(\cT_t = (\vu_i,\mA^\top\vu_i,\vv_i,\mA\vv_i)_{i=1}^t\).
    Then there exist matrices \(\mB_{11}\in\bbR^{t\times t}\), \(\mB_{12}\in\bbR^{t\times(n-t)}\), \(\mB_{21}\in\bbR^{(n-t)\times t}\), and orthogonal matrices \(\mR,\mS\in\bbR^{n \times n}\), all constructed deterministically from \(\cT_t\), such~that
    \[
        \mR^\top\mA\mS = \bmat{\mB_{11} & \mB_{12} \\ \mB_{21} & \mG}.
    \]
    Conditioned on the transcript \(\cT_t\), the matrices \(\mB_{11}\), \(\mB_{12}\), and \(\mB_{21}\) are deterministic while \(\mG\in\bbR^{(n-t) \times (n-t)}\) has iid \(\cN(0,1)\) entries.
\end{theorem}
\begin{proof}
Without loss of generality, we assume that the left queries \(\vu_1,\ldots,\vu_t\) are orthonormal, as are the right queries \(\vv_1,\ldots,\vv_t\), and we collect these queries into matrices
    \[
        \mU_t \defeq \bmat{\vu_1 & \cdots & \vu_t}
        \quad\text{and}\quad
        \mV_t \defeq \bmat{\vv_1 & \cdots & \vv_t}.
    \]
    We can also define \(\mU_{t,\perp}\) to be an orthonormal completion of \(\mU_t\), meaning that \(\mR_t = [\mU_t ~ \mU_{t,\perp}]\) is orthogonal.
    Similarly define \(\mV_{t,\perp}\) to be the orthonormal completion of \(\mV_t\).

    We start by showing at every timestep \(i \leq t\) that \(\mU_{i,\perp}^\top\mA\mV_{i,\perp}\in\bbR^{(n-i) \times (n-i)}\) is distributed as a Gaussian matrix, even when we condition on observing \(\cT_i\).
    We show this by induction, where the claim is trivial for \(i=0\).
    Then, we assume that the claim holds for timestep \(i\) and prove it for \(i+1\).

    Condition on the transcript \(\cT_i\).
    Recall that if two vectors are jointly Gaussian but uncorrelated, then they are independent.
    Hence, since \(\vu_{i+1}\) and \(\vv_{i+1}\) are orthogonal to \(\mU_{i+1,\perp}\) and \(\mV_{i+1,\perp}\), and by the rotational invariance of Gaussians, we know that
    \[
        \mU_{i,\perp}^\top\mA\mV_{i,\perp}
        \overset{\mathrm{dist}}{=}
        \bmat{\vu_{i+1} & \mU_{i+1,\perp}}^\top \mA \bmat{\vv_{i+1} & \mV_{i+1,\perp}}
    \]
    is distributed as a Gaussian matrix.
    We expand this Gaussian matrix as a block matrix:
    \[
        \bmat{\vu_{i+1} & \mU_{i+1,\perp}}^\top \mA \bmat{\vv_{i+1} & \mV_{i+1,\perp}}
        = \bmat{
            g_{11} & \vg_{12}^\top \\ \vg_{21} & \mG_{22}
        }.
    \]
    By the induction hypothesis, we know that \(\mU_{i,\perp}^\top\mA\mV_{i,\perp}\) is also independent of the transcript \(\cT_{i}\).
    The decomposition above further tells us that \(\mU_{i+1,\perp}^\top\mA\mV_{i+1,\perp}\) is distributed as a Gaussian matrix independent of \(g_{11}\), \(\vg_{12}\), and \(\vg_{21}\).
Since the output \((\mA^\top\vu_{i+1},\mA\vv_{i+1})\) is completely determined by the previous transcript \(\cT_{i}\) and the quantities \(g_{11}\), \(\vg_{12}\), and \(\vg_{21}\), we conclude that \(\mU_{i+1,\perp}^\top\mA\mV_{i+1,\perp}\) is independent of the updated transcript \(\cT_{i+1}\).

    With this fact in place, we can now construct the orthogonal matrices that give the promised block matrix decomposition.
    We let \(\mR \defeq [\mU_t ~ \mU_{t,\perp}]\) and \(\mS \defeq [\mV_t ~ \mV_{t,\perp}]\) so that we have
    \[
        \mR^\top\mA\mS
        = \bmat{
            \mU_t^\top\mA\mV_t & \mU_t^\top\mA\mV_{t,\perp} \\
            \mU_{t,\perp}^\top\mA\mV_t & \mU_{t,\perp}^\top\mA\mV_{t,\perp}
        }
        \overset{\mathrm{dist}}=
        \bmat{
            \mB_{11} & \mB_{12} \\
            \mB_{21} & \mG
        }.
    \]
    We see that \(\mB_{11} = \mU_t^\top\mA\mV_t\), \(\mB_{12}=\mU_t^\top\mA\mV_{t,\perp}\), and \(\mB_{21}=\mU_{t,\perp}^\top\mA\mV_{t}\).
    All three are computable deterministically since the transcript \(\cT_t\) identifies the matrices \(\mU_t^\top\mA\) and \(\mA\mV_t\) directly.
    This concludes the proof.
\end{proof}

\begin{lemma}[Control from the hidden Gaussian]\label{lem:hidden-gaussian}
    Let \(\mA\in\bbR^{n \times n}\) have iid \(\cN(0,1)\) entries, and fix any \(\mDelta\in\bbR^{n \times n}\).
    Suppose an algorithm computes \(t\) paired matvecs with \(\widehat\mA \defeq \mA+\mDelta\), and let \(r \defeq n-2t\).
    Then, conditioned on the transcript \(\cT_t\) of those matvecs, there exist
    independent random matrices \(\mH,\mF\in\bbR^{r \times r}\) such that \mH is Gaussian and
\[
        \norm{\mF}_2 \leq \norm{\mDelta}_2
\quad\text{and}\quad
        \norm{\widehat\mA^{-1}}_{\rm F}^2 \geq \sigma_{\min}(\mH+\mF)^{-2}
        \quad
        \text{almost surely.}
    \]
\end{lemma}
\begin{proof}
    This proof has two main ideas.
    First, we relate the minimum singular value of an \(n \times n\) matrix to its bottom right \((n-t) \times (n-t)\) Schur complement.
    Then, construct a series of rotations independent of most of the hidden Gaussian that reveals an \(n-2t\) dimensional submatrix of the Schur complement that depends only on the hidden Gaussian and the perturbation (i.e., not on the blocks \(\mB_{11}\), \(\mB_{12}\), and \(\mB_{21}\), which only have rank \(t\), resulting in the final \(n-2t\) dimensional matrices).

    Invoke the notation of the hidden Gaussian theorem (\cref{thm:hidden-gaussian}) to give orthogonal matrices \(\mR\) and \(\mS\).
    Defining \(\widehat\mDelta \defeq \mR^\top\mDelta\mS\), we look at the block matrix decomposition
    \[
        \mR^\top\widehat\mA\mS
        = \bmat{\mB_{11} & \mB_{12} \\ \mB_{21} & \mG}
        + \bmat{\widehat\mDelta_{11} & \widehat\mDelta_{12} \\ \widehat\mDelta_{21} & \widehat\mDelta_{22}}
        \eqqcolon \bmat{
            \mC_{11} & \mC_{12} \\ \mC_{21} & \mG + \widehat\mDelta_{22}
        }.
    \]
    Note that since \(\mA\) is Gaussian, the orthogonal matrices \(\mR\) and \(\mS\) do not change the rank of any objects, and we have that \(\mC_{11}\) is invertible.
    As such, we can expand the Schur complement of \(\mR^{\top}\widehat\mA\mS\):
    \[
        \mK \defeq \mG + \widehat\mDelta_{22} - \mL
        \quad
        \text{where } \mL \defeq \mC_{21}\mC_{11}^{-1}\mC_{12}.
    \] 
    A standard fact about Schur complements is that \(\sigma_{\min}(\mR^\top\widehat\mA\mS) \leq \sigma_{\min}(\mK)\).
    Therefore, we can bound
    \(
        \sigma_{\min}(\widehat\mA)
\leq \sigma_{\min}(\mK)
    \).
    Our goal is now to bound the minimum singular value of the Schur complement.

    Since \(\mC_{11}\in\bbR^{t \times t}\), we know that \(\mL\) has rank at most \(t\).
    So, we can find a matrix \(\mQ_{\mL}\in\bbR^{(n-t)\times t}\) with orthonormal columns that span the range of \mL.
    Let \(\mQ_{\mL,\perp}\in\bbR^{(n-t)\times(n-2t)}\) orthogonally complete \(\mQ_{\mL}\), and define the orthogonal matrix \(\mQ = [\mQ_{\mL} ~ \mQ_{\mL,\perp}]\).
    Further expand the block matrix
    \[
        \mQ^\top\mK = \bmat{
            \mQ_{\mL}^\top\mK \\ \mQ_{\mL,\perp}^\top\mK
        }
        = \bmat{
            \mZ \\ \mQ_{\mL,\perp}^\top(\mG + \widehat\mDelta_{22})
        },
    \]
    where we have defined \(\mZ\defeq \mQ_{\mL}^\top\mK\in\bbR^{t \times (n-t)}\).
    Since this matrix is short and wide, there exists a subspace \(\cW\subseteq\bbR^{n-t}\) such that every \(\vw\in\cW\) has \(\mZ\vw=\vec0\).
    Let \(\mW\in\bbR^{(n-t)\times(n-2t)}\) be an orthonormal matrix spanning such a space, constructed deterministically by looking only at \mZ.
    Then,
\[
        \mQ^\top\mK\mW
        = \bmat{ \mZ\mW \\ \mQ_{\mL,\perp}^\top(\mG+\widehat\mDelta_{22})\mW }
        = \bmat{ \mat0 \\ \mH + \mF }
        \quad
        \text{where }
        \mH \defeq \mQ_{\mL,\perp}^\top\mG\mW
        \text{ and }
        \mF \defeq \mQ_{\mL,\perp}^\top\widehat\mDelta_{22}\mW.
    \]
    Next, we confirm that \mH is distributed as a Gaussian matrix.
    To do so, we condition on the transcript \(\cT_t\) and the realization of \(\mZ\), so that \mW, \(\mQ_{\mL}\), \(\mQ_{\mL,\perp}\), and \(\widehat\mDelta_{22}\) are all fixed matrices.
    Next, we know that \(\mQ_{\mL}^\top\mG\) and \(\mQ_{\mL,\perp}^\top\mG\) are independent Gaussian matrices.
    Since \mW depends only on \mZ, and since \mZ depends on \(\mQ_{\mL}^\top\mG\) but not \(\mQ_{\mL,\perp}^\top\mG\), we know that \(\mW\) is independent of \(\mQ_{\mL,\perp}^\top\mG\).
    Therefore, our matrix \(\mH = (\mQ_{\mL,\perp}^\top\mG)\mW\) is the product of a Gaussian matrix and an independent orthogonal matrix.
    That is, \mH is Gaussian when conditioned on \(\cT_t\) and \mZ.
    Hence, \mH is still Gaussian when conditioned only on \(\cT_t\).
    
    The same logic shows that \mF remains independent of \mH.
We further have \(\norm{\mF}_2 \leq \norm{\widehat\mDelta_{22}}_2
    \leq \norm{\mDelta}_2\).
    So,
    \[
        \sigma_{\min}(\widehat\mA)
        \leq \sigma_{\min}(\mK)
        \leq \sigma_{\min}(\mK\mW)
        = \sigma_{\min}(\mQ^\top\mK\mW)
        = \sigma_{\min}(\mH+\mF).
    \]
    We finally conclude by observing
    \[
        \norm{\widehat\mA^{-1}}_{\rm F}^2
        \geq \frac1{\sigma_{\min}(\widehat\mA)^2}
        \geq \frac1{\sigma_{\min}(\mH+\mF)^2}.
    \]
\end{proof}

\paragraph{Reduction to estimating the Frobenius norm of the inverse.} The final stage of the argument involves showing that if we could perform linear solves with the matrix $\widehat \mA$ then we could approximate the Frobenius norm $\norm{\widehat\mA^{-1}}_{\rm F}$ by trace estimation. The following lemma shows that because of the hidden Gaussian theorem, after $\Omega(n)$ queries the value of this quantity is still inapproximable with constant probability. This is where we use \cref{thm:gaussian-near-singularity} to show that the hidden Gaussian matrix forces a long tail in the distribution of $\norm{\widehat\mA^{-1}}_{\rm F}$.
\begin{lemma}[A lower bound for inverse Frobenius estimation]
\label{lem:gaussian-factor-two-estimation}
Let \(\mA\in\bbR^{n\times n}\)
have iid \(\cN(0,1)\) entries, let \(\mDelta\in\bbR^{n\times n}\)
be independent of \(\mA\) with
\(\norm{\mDelta}_2\le\sqrt n/3\) almost surely, and set
\(\widehat\mA\defeq\mA+\mDelta\).
There is a universal constant $\rho>0$ such that every randomized estimator \(\widehat F\) using
at most \(t\le3n/32\) paired matvecs satisfies
\[  
        \widehat F\notin
        \left[\,
            \frac12\norm{\widehat\mA^{-1}}_{\rm F}^2,\ 
            2\norm{\widehat\mA^{-1}}_{\rm F}^2\,
        \right]
    \quad\text{with prob. }\ge\rho.
\]
\end{lemma}

\begin{proof}

By the hidden Gaussian construction (\cref{thm:hidden-gaussian} and \cref{lem:hidden-gaussian}),
conditioning first on \(\mDelta\) if it is random, conditional on the transcript
\(\cT_t\), there are a standard Gaussian matrix
\(\mH\in\bbR^{r\times r}\) and a possibly random matrix
\(\mF\in\bbR^{r\times r}\), independent of \(\mH\) given \(\cT_t\), satisfying
\[
    r=n-2t\ge\frac n2,
    \qquad
    \norm{\mF}_2\le\norm{\mDelta}_2
        \le\frac{\sqrt n}{3}\le\frac{\sqrt r}{2},
    \qquad
    \norm{\widehat\mA^{-1}}_{\rm F}^2
        \ge\sigma_{\min}(\mH+\mF)^{-2}.
\]
Let \(C\ge 1\) be an absolute constant to be chosen later, and set \(s=4\sqrt{C}\). \cref{thm:gaussian-near-singularity}, applied
to $\mH+\mF$ conditionally on \(\cT_t\) and \(\mF\), and then averaged
over \(\mF\), gives an absolute
constant \(p>0\) such that
\[
    \bbP\!\left[
        \sigma_{\min}(\mH+\mF)
        \le\frac{1}{s\sqrt r}
        \,\middle|\,\cT_t
    \right]\ge p.
\]
Since \(s^2r=16Cr\ge8Cn\), it follows that
\begin{equation}\label{eq:gaussian-factor-two-conditional-tail}
    \bbP\!\left[
        \norm{\widehat\mA^{-1}}_{\rm F}^2\ge8Cn
        \,\middle|\,\cT_t
    \right]\ge p.
\end{equation}
Let \(\cE=\{\widehat F\le2Cn\}\) denote the event where the estimate $\widehat F$ is not too large. If $\cE$ holds with probability $\ge 1/2$, then
\cref{eq:gaussian-factor-two-conditional-tail} gives
\begin{align*}
    \bbP\!\left[
        \widehat F<\frac12\norm{\widehat\mA^{-1}}_{\rm F}^2
    \right]
    &\ge\bbP\!\left[
        \cE\text{ and }\big\{
            \norm{\widehat\mA^{-1}}_{\rm F}^2\ge8Cn
        \big\}
    \right] 
    \ge p\cdot\bbP[\cE]\ge\frac{p}{2}.\end{align*}
Otherwise, if \(\bbP[\cE]<1/2\), then 
\begin{align}
    \bbP\!\left[
        \widehat F>2\norm{\widehat\mA^{-1}}_{\rm F}^2
    \right]
    &\ge\bbP\!\left[
        \neg\cE\text{ and }\big\{
            \norm{\widehat\mA^{-1}}_{\rm F}^2\le Cn
        \big\}
    \right] 
    \ge\bbP\!\left[
        \norm{\widehat\mA^{-1}}_{\rm F}^2\le Cn
    \right]-\frac12.\label{eq:frobenius-upper}
\end{align}
It now suffices to show that with high probability $\norm{\widehat\mA^{-1}}_{\rm F}^2$ cannot be too large.
By \cite[Lemma~3.2]{sankar2006smoothed}, applied with Gaussian
variance one and deterministic shift \(\mDelta\), for every fixed
unit vector \(\vv\in\bbR^n\) and $u>0$,
\begin{equation}\label{eq:gaussian-inverse-directional-tail}
    \bbP_{\mA}\!\left[
        \norm{\widehat\mA^{-1}\vv}_2>u
    \right]
    \le\frac{\sqrt{2/\pi}}{u}.
\end{equation}

Now take an independent \(\vg\sim\cN(\vec0,\mI_n)\) and put
\(Y=\norm{\widehat\mA^{-1}\vg}_2^2 = \vg^\top(\widehat\mA\widehat\mA^\top)^{-1}\vg\). Conditional on
\(\widehat\mA\), standard trace estimation bounds yield:
\[
    \E[Y\mid\widehat\mA]
    =\norm{\widehat\mA^{-1}}_{\rm F}^2,
    \qquad
    \E[Y^2\mid\widehat\mA]
    \le3\norm{\widehat\mA^{-1}}_{\rm F}^4.
\]
Therefore, applying the Paley--Zygmund inequality gives
\[
    \bbP\!\left[
        Y\ge\frac12\norm{\widehat\mA^{-1}}_{\rm F}^2
        \,\middle|\,\widehat\mA
    \right]
    \ge\frac1{12}.
\]
On the other hand, conditioning on \(\vg\) and using
\cref{eq:gaussian-inverse-directional-tail},
\[
    \bbP[Y>Cn/2]
    \le
    \sqrt{\frac2\pi}\,
    \frac{\E\norm{\vg}_2}{\sqrt{C n/2}}
    \le\frac2{\sqrt{\pi C}}.
\]
The event \(\norm{\widehat\mA^{-1}}_{\rm F}^2> Cn\),
together with
\(Y\ge\norm{\widehat\mA^{-1}}_{\rm F}^2/2\),
implies \(Y> Cn/2\), so
\[
    \bbP\!\left[
        \norm{\widehat\mA^{-1}}_{\rm F}^2 > C n
    \right]
    \le12\bbP[Y>Cn/2]
    \le\frac{24}{\sqrt{\pi C}}.
\]
Setting \(C=10^4\) makes the last bound less than \(1/4\), implying that $\bbP\big[\norm{\widehat\mA^{-1}}_{\rm F}^2 \le C n\big]\ge 3/4$. Putting this into \cref{eq:frobenius-upper}, we get $\bbP\big[
        \widehat F>2\norm{\widehat\mA^{-1}}_{\rm F}^2
    \big]\ge 1/4$, which gives the claim.
\end{proof}

We are now ready to give the proof of \cref{thm:smoothed-lower-bound}, which involves a straightforward reduction from inverse Frobenius estimation to linear system solving.

\begin{proof}[Proof of \cref{thm:smoothed-lower-bound}]
Draw \(\vg_1,\ldots,\vg_m\) independently from
\(\cN(\vec0,\mI_n)\).
Run the solver on the same matrix \(\widehat\mA\) with
right-hand sides \(\vg_i\) and fresh independent seeds,
also independent of the matrices and right-hand sides.
Let \(\widetilde{\vx}_i\) be the resulting outputs, and define
\[
    F\defeq\norm{\widehat\mA^{-1}}_{\rm F}^2,
    \qquad
    T\defeq\frac1m\sum_{i=1}^m
        \norm{\widehat\mA^{-1}\vg_i}_2^2,
    \qquad
    \widehat F\defeq\frac{10}{9m}\sum_{i=1}^m
        \norm{\widetilde{\vx}_i}_2^2.
\]
Each solve fails with probability at most \(1-p\), so the
probability that any solve fails is at most \(m(1-p)\). Now, proceeding via standard analysis of Hutchinson's trace estimator \cite{hutchinson1989stochastic}, we have
\[
    \bbP\!\left[T\notin[F/2,3F/2]\right]\le\frac8m.
\]
When every solve succeeds, we can ensure that $\widehat F$ is a good approximation of $T$, namely, $T\le\widehat F\le\frac43T$.
Combining this with the successful trace estimation, it follows that $\frac F2\le\widehat F\le2F$. A union bound gives the failure probability \(m(1-p)+8/m\). Choosing $m\ge 16/\rho$ and $p\le 1 - \rho^2/32$, we get \(m(1-p)+8/m\le \rho\), obtaining the claim from \cref{lem:gaussian-factor-two-estimation}.
\end{proof}

\subsection{Proof of \texorpdfstring{\cref{thm:gaussian-near-singularity}}{\cref*{thm:gaussian-near-singularity}}}
\label{sec:gaussian-near-singularity}

We conclude this section by giving an elementary proof of the fact that a Gaussian matrix remains ill-conditioned after an arbitrary perturbation of bounded size.
Letting $\widehat\mA=\mA+\mDelta$, our goal is to show that
$\norm{\widehat\mA^{-1}}_2\ge s\sqrt n$ with decent probability. Without loss of generality, assume that $\mDelta$ is deterministic, since we can integrate out over $\mDelta$ at the end.
The proof is based on two key ingredients. We start by showing that,
for any given direction $\vv$, the matrix $\widehat\mA^{-1}$
cannot be too small in that direction, i.e., that
$\norm{\widehat\mA^{-1}\vv}_2$ is bounded away from zero with
decent probability.
This strategy mirrors the one used by \cite{sankar2006smoothed},
who seek to show that $\norm{\widehat\mA^{-1}}_2$ cannot be too
large (the opposite direction). And in fact, the following lemma
mirrors their Lemma 3.2, with our lower bound matching their upper
bound up to constant factors. However, as this is an
anti-concentration result, the proof is quite different.

\begin{lemma}[Lower bound]\label{lem:gaussian-reverse}
Let $\widehat\mA=\mA+\mDelta$ be defined as in
\cref{thm:gaussian-near-singularity}.
There is an absolute constant $c>0$ such that for every
deterministic unit vector $\vv\in\bbR^n$ and every $t\ge1$,
\begin{equation}\label{eq:gaussian-reverse}
    \bbP[\norm{\widehat\mA^{-1}\vv}_2\ge t]\ge\frac ct.
\end{equation}
\end{lemma}

Note that the above lower bound also holds if we let $\vv$ be a
uniformly random unit vector independent of $(\mA,\mDelta)$, which
is how we will use it in our analysis. However, by itself,
\cref{lem:gaussian-reverse} is not sufficient for
\cref{thm:gaussian-near-singularity}, as it does not match
the correct dimension dependence of $\sqrt n$.
To address this, the second ingredient in our argument is that,
for a random unit $\vv$, the quantity
$\norm{\widehat\mA^{-1}\vv}_2$ cannot be large while at the same
time $\norm{\widehat\mA^{-1}}_2$ is small, i.e., we show a sharper
upper bound for the probability of the intersection of these
two events.

\begin{lemma}[Upper bound]\label{lem:gaussian-intersection}
Let $\widehat\mA=\mA+\mDelta$ be defined as in
\cref{thm:gaussian-near-singularity}, and let
$\vv\in \bbS^{n-1}$ be a uniformly random unit vector independent
of $(\mA,\mDelta)$. For every $s,t\ge1$,
\begin{align*}
    \bbP\Big[
        \norm{\widehat\mA^{-1}\vv}_2\ge t
        \quad\text{and}\quad
        \norm{\widehat\mA^{-1}}_2\le s\sqrt n
    \Big]
    \le\frac{3s^2}{t^2}.
\end{align*}
\end{lemma}

\cref{thm:gaussian-near-singularity} now follows by
combining these two probability estimates.

\begin{proof}[Proof of \cref{thm:gaussian-near-singularity}]
Define our target event
$\cE=\{\norm{\widehat\mA^{-1}}_2\ge s\sqrt n\}$.
Combining the upper and lower estimates from
\cref{lem:gaussian-reverse,lem:gaussian-intersection}, we obtain that for a uniformly
random unit vector $\vv\in \bbS^{n-1}$ and $t\ge1$,
\begin{align*}
    \frac ct
    &\le\bbP\big[\norm{\widehat\mA^{-1}\vv}_2\ge t\big] \\
    &\le\bbP\Big[
        \norm{\widehat\mA^{-1}\vv}_2\ge t
        \ \text{and}\ \neg\cE
    \Big]+\bbP[\cE] \\
    &\le\frac{3s^2}{t^2}+\bbP[\cE].
\end{align*}
By choosing sufficiently large $t=\cO(s^2)$, we obtain
$\bbP[\cE]=\Omega(1/s^2)$.
Since
\[
    \cE=\left\{
        \sigma_{\min}(\widehat\mA)\le\frac1{s\sqrt n}
    \right\},
\]
this proves the desired bound for a fixed $\mDelta$. Averaging over $\mDelta$ concludes the proof.
\end{proof}

\paragraph{Auxiliary lemma.}
In the proofs of \cref{lem:gaussian-reverse,lem:gaussian-intersection}, we will repeatedly use that
centering a Gaussian vector increases the expectation of a
decreasing function of its squared norm, including after a
deterministic linear transformation. This is a consequence of
Anderson's inequality \cite{anderson1955integral}, see
Lemma 28.9 in \cite{polyanskiy2025information}.

\begin{importedlemma}[Anderson's inequality \cite{anderson1955integral}]
\label{lem:gaussian-centering}
If $\vz$ is standard Gaussian, $\mT$ and $\vm$ are deterministic,
and $f:[0,\infty)\to[0,\infty]$ is nonincreasing and measurable,
then
\begin{equation}
    \E\left[f\!\left(\norm{\mT(\vz+\vm)}_2^2\right)\right]
    \le
    \E\left[f\!\left(\norm{\mT\vz}_2^2\right)\right].
\end{equation}
\end{importedlemma}

\subsubsection{Proof of \texorpdfstring{\cref{lem:gaussian-reverse}}{\cref*{lem:gaussian-reverse}}:
A small-ball estimate}

We start by rewriting $\norm{\widehat\mA^{-1}\vv}_2$ similarly
as in the proof of Lemma 3.2 in \cite{sankar2006smoothed}.
Let $\widehat\mA=\mA+\mDelta\in\bbR^{n\times n}$ be as in
\cref{thm:gaussian-near-singularity} and fix a unit
vector $\vv$. Choose an orthogonal matrix $\mQ$ such that
$\mQ\vv=\ve_1$, and write the rows of $\mQ\widehat\mA$ as
$\vb_1^\top,\ldots,\vb_n^\top$.
Let $\vu$ be a unit vector orthogonal to the span of
$\vb_2,\ldots,\vb_n$.
The vector
$\widehat\mA^{-1}\vv=(\mQ\widehat\mA)^{-1}\ve_1$
is orthogonal to these rows and has inner product one with
$\vb_1$. Consequently,
\begin{equation}\label{eq:gaussian-sst}
    \norm{\widehat\mA^{-1}\vv}_2
    =\frac1{|\langle\vb_1,\vu\rangle|}
    =\frac1{|Z+\mu|},
    \qquad
    \mu=\langle\vr,\vu\rangle,
\end{equation}
where $\vr=\E[\vb_1]$ and $Z$ is a standard Gaussian
independent of $\mu$.

The next lemma controls the conditional mean $\mu$ in this
representation.

\begin{lemma}[A normal to Gaussian rows]
\label{lem:gaussian-normal}
Let $\mB\in\bbR^{(n-1)\times n}$ be standard Gaussian plus a
deterministic matrix of norm at most $a$.
Let $\vu$ be a unit vector orthogonal to the row vectors of
$\mB$. If $n>2a^2+2$, then, for every deterministic
$\vw\in\bbR^n$,
\begin{equation}\label{eq:gaussian-normal}
    \E\bigl[|\langle\vw,\vu\rangle|^2\bigr]
    \le\frac{2\norm{\vw}_2^2}{n-2a^2-2}.
\end{equation}
\end{lemma}

\begin{proof}
We will prove this via an induction over $n$.
Note that it suffices to show this for $\vw=\ve_1$ and every
deterministic shift, since for any other $\vw$ we can simply
rotate it back to $\ve_1$ and apply the same rotation
to the deterministic shift.

If $0<n-2a^2-2\le2$, the claim follows from $u_1^2\le1$,
so suppose that $n-2a^2-2>2$, and write
$\mB=[\vb\ \mC]$.
Since $\mB\vu=\vec0$, vector $\vu$ is proportional to
$(1,-(\mC^{-1}\vb)^\top)^\top$, and so
\[
    u_1^2=\frac1{1+\norm{\mC^{-1}\vb}_2^2}.
\]
Conditional on $\mC$, the column $\vb$ is a standard Gaussian
vector plus a deterministic mean.
Applying \cref{lem:gaussian-centering} with
$f(x)=(1+x)^{-1}$ we get
\begin{equation}\label{eq:gaussian-normal-center}
    \E[u_1^2]
    \le\E\left[\frac1{1+\norm{\mC^{-1}\vh}_2^2}\right],
\end{equation}
where $\vh\sim\cN(\vec0,\mI_{n-1})$ is independent of $\mC$.

Writing $\vh=R\vq$ for unit length $\vq$,
\cref{eq:gaussian-sst} for the $(n-1)\times(n-1)$ matrix
$\mC$ yields
\[
    \norm{\mC^{-1}\vh}_2=\frac R{|Z+\nu|},
\]
where $\nu=\langle\vr,\vu'\rangle$.
Here, conditioned on $\vq$, $\vr$ is deterministic with
$\norm{\vr}_2\le a$, and $\vu'$ is a unit vector orthogonal
to $n-2$ Gaussian rows in $\bbR^{n-1}$ shifted by a
deterministic matrix with norm at most $a$.
The induction hypothesis gives
\[
    \E[\nu^2]
    =\E\bigl[\E[\nu^2\mid\vq]\bigr]
    \le\frac{2a^2}{n-2a^2-3}.
\]
Since $R\sim\chi_{n-1}$ is independent of $(Z,\nu)$ and
$Z$ is independent of $\nu$,
\cref{eq:gaussian-normal-center} yields
\begin{align*}
    \E[u_1^2]
    &\le\E\left[\frac{(Z+\nu)^2}{R^2}\right]
    =\frac{1+\E[\nu^2]}{n-3}
    \le\frac{1+2a^2/(n-2a^2-3)}{n-3}
    \le\frac2{n-2a^2-2},
\end{align*}
where we used an inverse chi-squared moment formula
$\E[1/R^2]=1/(n-3)$
(e.g., Lemma A.8 in \cite{tropp2011structure}).
\end{proof}

We are now ready to complete the proof of
\cref{lem:gaussian-reverse}.

\begin{proof}[Proof of \cref{lem:gaussian-reverse}]
Writing $\norm{\widehat\mA^{-1}\vv}_2$ in the form of
\cref{eq:gaussian-sst} and applying
\cref{lem:gaussian-normal} with $a=\sqrt n/2$ gives
\[
    \E[\mu^2]
    \le\frac{2a^2}{n-2a^2-2}
    =\frac n{n-4}
    \le2.
\]
Markov's inequality therefore implies
$\bbP[|\mu|\le2]\ge1/2$.
On this event, the interval $[-\mu-1/t,-\mu+1/t]$ has
length $2/t$ and is contained in $[-3,3]$.
If $\phi$ denotes the standard Gaussian density,
independence of $Z$ and $\mu$ gives
\[
    \bbP[\norm{\widehat\mA^{-1}\vv}_2\ge t]
    =\bbP[|Z+\mu|\le1/t]
    \ge\frac12\,\frac{2\phi(3)}t
    \ge\frac1{256t}.
\]
\end{proof}

\subsubsection{Proof of \cref{lem:gaussian-intersection}:
Inverse spectral tail estimate}

Recall that our goal is to bound the probability of the
intersection of events
$\norm{\widehat\mA^{-1}\vv}_2\ge t$ and
$\norm{\widehat\mA^{-1}}_2\le s\sqrt n$.
We will use the latter event to control the two smallest
singular values of $\widehat\mA$.
In order to control the remaining ones, we use the following
tail estimate which relies on the inverse chi-squared moment
with three degrees of freedom.

\begin{lemma}[Inverse spectral tail]\label{lem:gaussian-bulk}
Let $\widehat\mA=\mA+\mDelta$ be defined as in
\cref{thm:gaussian-near-singularity}. Then
\begin{equation}\label{eq:gaussian-bulk}
    \E\left[\sum_{j=1}^{n-2}\sigma_j(\widehat\mA)^{-2}\right]
    \le n-2.
\end{equation}
\end{lemma}

\begin{proof}
Delete the last two columns of $\widehat\mA$ to obtain
$\mC=[\vc_1,\ldots,\vc_{n-2}]\in\bbR^{n\times(n-2)}$.
For each $j$, let $\mB_j$ consist of the columns of $\mC$
other than $\vc_j$, and let $\mP_j$ be the orthogonal
projector onto their span.
A diagonal Schur complement gives
\begin{align*}
    \bigl[(\mC^\top\mC)^{-1}\bigr]_{jj}
    &=
    \left(
        \vc_j^\top\vc_j
        -\vc_j^\top\mB_j
        (\mB_j^\top\mB_j)^{-1}
        \mB_j^\top\vc_j
    \right)^{-1}
    =\norm{(\mI-\mP_j)\vc_j}_2^{-2}.
\end{align*}
Summing over $j$ yields the column-distance identity for the
trace of $(\mC^\top\mC)^{-1}$:
\begin{equation}\label{eq:gaussian-distance}
    \tr\bigl((\mC^\top\mC)^{-1}\bigr)
    =\sum_{j=1}^{n-2}
    \operatorname{dist}\!\left(
        \vc_j,\Span\{\vc_i:i\ne j\}
    \right)^{-2}.
\end{equation}

Condition on all columns other than $\vc_j$.
Their span has dimension $n-3$ almost surely.
In an orthonormal basis for its orthogonal complement,
$(\mI-\mP_j)\vc_j$ is a standard Gaussian vector in
$\bbR^3$ plus a fixed mean.
Therefore, by \cref{lem:gaussian-centering}, each
summand in \cref{eq:gaussian-distance} has conditional
expectation at most $\E[\chi_3^{-2}]=1$. Hence
\[
    \E\bigl[\tr\bigl((\mC^\top\mC)^{-1}\bigr)\bigr]\le n-2.
\]

Finally,
$\widehat\mA\widehat\mA^\top\succeq\mC\mC^\top$.
Monotonicity of ordered eigenvalues gives
$\lambda_i(\widehat\mA\widehat\mA^\top)
\ge\lambda_i(\mC\mC^\top)$,
and since $\mC\mC^\top$ has two zero eigenvalues,
it follows that
\[
    \sum_{j=1}^{n-2}\sigma_j(\widehat\mA)^{-2}
    \le\tr\bigl((\mC^\top\mC)^{-1}\bigr).
\]
Taking expectations proves \cref{eq:gaussian-bulk}.
\end{proof}

Here, it was crucial that we omitted two smallest singular
values of $\widehat\mA$, as opposed to one.
Otherwise, the inverse moment of the chi-squared distribution
with 2 degrees of freedom would be infinite,
$\E[\chi_2^{-2}]=\infty$, and the approach does not give a
useful upper bound on
$\E\bigl[\sum_{j=1}^{n-1}\sigma_j(\widehat\mA)^{-2}\bigr]$.

We are now ready to prove
\cref{lem:gaussian-intersection}.

\begin{proof}[Proof of \cref{lem:gaussian-intersection}]
Recall that $\vv$ is a uniformly random unit vector from
$\bbS^{n-1}$, so $\E_{\vv}[\vv\vv^\top]=\mI_n/n$.
Letting
$\cE=\{\norm{\widehat\mA^{-1}}_2\ge s\sqrt n\}$
and applying \cref{lem:gaussian-bulk}, we have
\begin{align*}
    \E\bigl[
        \norm{\widehat\mA^{-1}\vv}_2^2
        \ \mathbbm1_{\neg\cE}
    \bigr]
    &=
    \frac1n\E\bigl[
        \norm{\widehat\mA^{-1}}_{\rm F}^2
        \ \mathbbm1_{\neg\cE}
    \bigr] \\
    &\le 2s^2
    +\frac1n\E\left[\sum_{j=1}^{n-2}
        \sigma_j(\widehat\mA)^{-2}\right]
    \le3s^2,
\end{align*}
where we use that $\sigma_n(\widehat\mA)^{-2}$ and
$\sigma_{n-1}(\widehat\mA)^{-2}$ are each at most $s^2n$
when $\cE$ fails.
Applying Markov's inequality to the nonnegative random variable
$\norm{\widehat\mA^{-1}\vv}_2^2 \ \mathbbm1_{\neg\cE}$, we get
\[
    \bbP\big[
        \norm{\widehat\mA^{-1}\vv}_2\ge t
        \ \text{and}\ \neg\cE
    \big]
    \le\frac{3s^2}{t^2}.
\]
\end{proof}

\section{Lower bounds against deterministic algorithms}
\label{sec:det-lower-bounds}

In this section, we prove \cref{thm:det-lower-bound-to-krylov}, showing that no deterministic algorithm can solve a general linear system to small backward error faster than a Krylov method that solves the normal equations.
Recall from \cref{eq:t-ne} that \(t_{\rm ne}(\mA,\vb)\) denotes the minimum odd degree \(t=2k+1\) that suffices for the Krylov subspace \(\cK_{k}(\mA^\top\mA,\mA^\top\vb)\) to contain a vector \(\vx\) which has small backward error \(\mathrm{berr}(\mA,\vb,\vx)\leq\eps\).
It takes \(2k+1\) matvecs with \mA to construct the Krylov subspace \(\cK_{k}(\mA^\top\mA,\mA^\top\vb)\), so \(t_{\rm ne}\) is counting matvecs while \(k\) is measuring the depth of the Krylov subspace.

We then consider an arbitrary deterministic matvec algorithm \(\Phi(\mA,\vb)\), which is guaranteed to return a vector \(\widetilde\vx\) of backward error at most \(\eps\) on all inputs.
We let \(t_{\Phi}(\mA,\vb)\) denote the number of matvecs that \(\Phi\) exerts on the input \((\mA,\vb)\).
Our goal is to show that
\[
    \sup_{\mA\in\cF_{\kappa}} t_{\Phi}(\mA,\vb) \geq \frac14\left(\sup_{\mA\in\cF_{\kappa}} t_{\rm ne}(\mA,\vb) - 1 \right),
\]
where \(\cF_{\kappa}\) is the set of all \(n\times n\) matrices with condition number at most \(\kappa\).

We will prove something much more general, which recovers this guarantee in the special case that we focus on backward error over matrices of bounded condition number.
Our lower bound will apply to any context in which we may ``test'' if we have sufficiently solved our linear algebra problem, and if our problem instances are orthogonally invariant.
\begin{setting}[General Inverse Problem]
    \label{setting:general-inverse-problem}
    Fix a set \(\cF\subseteq\bbR^{n \times n}\) of invertible matrices and a test function \(\textsc{acc}:\bbR^{n \times n} \times \bbR^n \times \bbR^n \to \{0,1\}\).
We assume the following relations hold for all \(\mA\in\cF\), all orthogonal matrices \(\mU,\mV\in\bbR^{n \times n}\), and all vectors \(\vb,\vz\in\bbR^n\):
    \begin{enumerate}
        \item \textbf{Orthogonal invariance of \cF:~} We have \(\mU^\top\mA\mV\in\cF\).
        \item \textbf{Orthogonal invariance of \textsc{acc}:~} If \(\mU\vb=\vb\), then \(\textsc{acc}(\mU^\top\mA\mV,\vb,\vz) = \textsc{acc}(\mA,\vb,\mV\vz)\).
        \item \textbf{Solvability:~} We always have \(\textsc{acc}(\mA,\vb,\mA^{-1}\vb)=1\).
    \end{enumerate}
\end{setting}

For instance, the backward-error test at level \(\eps\) is
\begin{equation}
    \textsc{acc}_{\rm berr}(\mA,\vb,\vz) = \mathbbm1_{\{\mathrm{berr}(\mA,\vb,\vz) \leq \eps\}},
    \label{eq:acc-berr}
\end{equation}
and the pair \((\cF_{\kappa},\textsc{acc}_{\rm berr}(\mA,\vb,\vz))\) satisfies the requirements of \cref{setting:general-inverse-problem}.
Most crucially, the orthogonal invariance of this test function follows from
\[
    \mathrm{berr}(\mU^\top\mA\mV,\vb,\vz)
    = \frac{\norm{\mU^\top\mA\mV\vz - \vb}_2}{\norm{\mA}_2 \norm{\vz}_2}
    = \frac{\norm{\mA\mV\vz - \vb}_2}{\norm{\mA}_2 \norm{\mV\vz}_2}
    = \mathrm{berr}(\mA,\vb,\mV\vz).
\]
Our core technical result is that a Krylov method is always a near-optimal deterministic matvec algorithm when \cref{setting:general-inverse-problem} applies.
In particular, we define the \emph{right-side Krylov subspace} \(\cR_{k}(\mA,\vb) \defeq \cK_{k}(\mA^\top\mA,\mA^\top\vb)\), and let \(t_{\rm k\text{-}opt}(\mA,\vb)\) be the odd degree \(2k+1\) needed for \(\cR_{k}(\mA,\vb)\) to contain a vector \(\vx\) that passes the test \(\textsc{acc}(\mA,\vb,\vx)=1\).
That is,
\begin{equation}
    \label{eq:t-k-opt}
    t_{\rm k\text{-}opt}(\mA,\vb) \defeq \min\left\{
        2k+1
        ~:~
        k \geq 0,\quad \exists \vx\in\cR_{k}(\mA,\vb) \text{ such that } \textsc{acc}(\mA,\vb,\vx) = 1
    \right\}.
\end{equation}
We can think of \(t_{\rm k\text{-}opt}\) as an idealized Krylov method, as it counts the number of matvecs needed to build a Krylov subspace for the normal equations that contains a valid solution.
We say the method is idealized because without further information, there is no guarantee that we can efficiently determine if there exists \(\vx\in\cR_{k}(\mA,\vb)\) that passes the test.
Nonetheless, as we will show later on, the parameter \(t_{\rm k\text{-}opt}\) is a valuable quantity for lower bounds because it cleanly relates deterministic matvec complexity to questions about polynomial approximation.
Our main technical result is that building a Krylov subspace for the normal equations is a near-optimal deterministic strategy:

\begin{theorem}[Optimality of Krylov on the normal equations]
    \label{thm:krylov-opt}
    Instate the notation and assumptions of \cref{setting:general-inverse-problem}.
    Fix any vector \(\vb\in\bbR^n\).
    Consider any deterministic matvec algorithm \(\Phi(\mA,\vb)\), and suppose that for all \(\mA\in\cF\) we have \(\textsc{acc}(\mA,\vb,\Phi(\mA,\vb))=1\).
    Let \(t_{\Phi}(\mA,\vb)\) denote the number of matvecs \(\Phi\) expends in this process.
    Then,
    \[
        \sup_{\mA\in\cF} t_{\Phi}(\mA,\vb)
        \geq
        \frac14 \left(
            \sup_{\mA\in\cF} t_{\rm k\text{-}opt}(\mA,\vb) - 1
        \right).
\]
\end{theorem}

To prove this result, we must engage with the fact that \cF may contain asymmetric matrices.
In such cases, a matvec algorithm at any step can decide to compute the matvecs \(\vx\mapsto\mA\vx\) or \(\vy\mapsto\mA^\top\vy\).
To simplify the analysis of such methods, we constrain our view to algorithms that always alternate between these two maps.

\begin{definition}[Alternating matvecs algorithm]
    We say that a matvec algorithm is \emph{alternating} if, at step \(i\), it performs the computations
    \[
        \begin{cases}
            \vy_i = \mA\vx_i & i \text{ is odd} \\
            \vx_i = \mA^\top\vy_i & i \text{ is even.}
        \end{cases}
    \]
\end{definition}

Note that any (perhaps non-alternating) matvec algorithm that uses \(t\) matvecs can be simulated by an alternating algorithm with \(2t\) matvecs.
This slightly moves the goalposts of the claim we seek to prove.
Indeed, we shall prove the following:

\begin{theorem}[Optimality of Krylov on the normal equations, alternating version]
    \label{thm:krylov-opt-alternating}
    Consider the same setup as \cref{thm:krylov-opt}, except that we further assume that \(\Phi\) is alternating.
    Then,
    \[
        \sup_{\mA\in\cF} t_{\Phi}(\mA,\vb)
        \geq
        \frac12 \left(
            \sup_{\mA\in\cF} t_{\rm k\text{-}opt}(\mA,\vb) - 1
        \right).
    \]
\end{theorem}

To prove this result, we will define a suitable family of rotation matrices that let us map one subspace of vectors to another subspace of vectors.

\begin{definition}[Nested rotations]
    Let \(\va_1,\ldots,\va_m\) and \(\vb_1,\ldots,\vb_m\) be vectors such that
    \(\cS_i = \Span\{\va_1,\ldots,\va_i\}\) and \(\cM_i = \Span\{\vb_1,\ldots,\vb_i\}\)
    have \(\dim\cS_i \leq \dim\cM_i\) for all \(1 \leq i \leq m\).
    For any orthogonal matrix \(\mQ\) with \(\mQ\cS_{m-1} \subseteq \cM_{m}\), we write \(\mR \geqq_m \mQ\) with domain \(\va_1,\ldots,\va_m\) and range \(\vb_1,\ldots,\vb_m\), if
    \begin{itemize}
        \item \textbf{The past is unchanged:~} \(\mR\vx = \mQ\vx\) for all \(\vx\in\cS_{m-1}\), and
        \item \textbf{Each new vector is mapped from domain to range:~} \(\mR\va_m\in\cM_m\).
    \end{itemize}
\end{definition}
The first item ensures a \emph{nesting property}, that \mR matches \mQ in the way it maps the first \(m-1\) elements of the domain to the first \(m-1\) elements of the range.
The second item ensures that \mR maps the \(m\)th item successfully from domain to range.
We can use the Gram-Schmidt procedure to construct nested rotations from any domain to any range.

Additionally, it will be helpful to think about \emph{transcripts}.
For a deterministic alternating matvec algorithm, the transcript after the first \(i\) matvecs is the vector \vb and the list of \((\vx_j, \vy_j)\) pairs:
\[
    \cT_i \defeq \bigl(\vb, (\vx_1,\vy_1), (\vx_2,\vy_2), \ldots, (\vx_i,\vy_i)\bigr).
\]
We can think of any deterministic alternating matvec algorithm \(\Phi\) as a sequence of functions \(\Phi_i\) that map \(\cT_{i-1}\) to the next query vector.
That is, at step \(i\), any deterministic alternating \(\Phi\) is uniquely described as performing
\[
    \begin{cases}
        \vy_i = \mA\vx_i & \text{ where } \vx_i = \Phi_i(\cT_{i-1}) \text{ for } i \text{ odd, and} \\
        \vx_i = \mA^\top\vy_i & \text{ where } \vy_i = \Phi_i(\cT_{i-1}) \text{ for } i \text{ even.}
    \end{cases}
\]

Lastly, we will need to talk about the \emph{grade} of a Krylov subspace, which is traditionally defined for an arbitrary Krylov subspace, though we will only apply it to \(\cR_k(\mA,\vb)\).
For a matrix \mA and a vector \vb, the grade is the maximum possible achievable dimension of a Krylov subspace of \mA initiated from the vector \vb.
For instance, if \vb is not orthogonal to any eigenvector of \mA then the grade will equal the rank of \mA.
In the other extreme, if \vb is an eigenvector of \mA, then the grade will be exactly 1.

\begin{definition}[Grade]
    For a matrix \(\mA\in\bbR^{n \times n}\) and vector \(\vb\in\bbR^n\), the grade of \mA with respect to \vb is \(d \defeq \dim\cR_n(\mA,\vb)\).
\end{definition}

We are now prepared to prove the core technical step of the proof, showing that any deterministic method which does not perform enough matvecs to exhaust the grade of a Krylov subspace must return a vector that (up to rotation of the input \emph{and} output) lies in a Krylov subspace.
The nature of these rotations is what dictates the formal requirements of orthogonal invariance in \cref{setting:general-inverse-problem}.

\begin{theorem}[Krylov methods can simulate deterministic methods]
    \label{thm:krylov-det-simulate}
Fix any \(\mA\in\bbR^{n \times n}\), \(\vb\in\bbR^n\), and deterministic alternating matvec algorithm \(\Phi\).
    Let \(d\) be the grade of \mA with respect to \vb.
    There exist orthogonal matrices \(\mU,\mV\in\bbR^{n \times n}\) such that \(\mU\vb=\vb\) and the following statement is true for \(\mB = \mU^\top\mA\mV\):
    if \(\vz=\Phi(\mB,\vb)\) is returned after computing \(k < d\) matvecs, then \(\mV\vz\in\cR_{k}(\mA,\vb)\).
\end{theorem}
\begin{proof}
    Consider the sequence of matvecs needed to construct the right Krylov subspaces \(\cR_k(\mA,\vb)\).
    Define \(\vl_0 \defeq \vb\), \(\vr_i \defeq \mA^\top\vl_{i-1}\), and \(\vl_i \defeq \mA\vr_i\).
    Note that
    \[
\Span\{\vr_1,\ldots,\vr_i\} = \cR_{i-1}(\mA,\vb).
    \]
    To begin, we will construct a transcript
    \[
        \cT_k = \bigl(\vb, (\vx_1, \vy_1), (\vx_2, \vy_2), \ldots, (\vx_k,\vy_k)\bigr)
    \]
    in an iterative manner.
This construction begins in the base case \(i=0\), where we have \(\cT_0 = (\vb)\) and define \(\mU_0=\mV_0=\mI_n\).
    We proceed iteratively over \(i\geq1\), creating the pair \((\vx_i,\vy_i)\) in a manner that depends on if \(i\) is odd or even.
    We will also preserve a pair of rotation matrices \(\mU_i,\mV_i\) at each step.
    \begin{itemize}
        \item 
        For any odd \(i\), the algorithm selects a query vector \(\vx_i=\Phi_i(\cT_{i-1})\).
        Let \(\mV_i\) be any matrix with \(\mV_i \geqq_i \mV_{i-1}\) on domain \(\vx_1,\ldots,\vx_i\) and range \(\vr_1,\ldots,\vr_i\).
Define
        \[
\widetilde{\vx}_i = \mV_i \vx_i,
            \qquad
            \widetilde{\vy}_i \defeq \mA\widetilde{\vx}_i,
            \quad
            \vy_i \defeq \mU_{i-1}^\top \widetilde{\vy}_i,
            \quad\text{and}\quad
            \mU_i = \mU_{i-1}.
        \]

        \item
For any even \(i\), the algorithm selects a query vector \(\vy_i = \Phi_i(\cT_{i-1})\).
Let \(\mU_i\) be any matrix with \(\mU_i \geqq_{i+1} \mU_{i-1}\) on domain \(\vb,\vy_1,\ldots,\vy_i\) and range \(\vl_0,\ldots,\vl_i\).
        Define
        \[
            \widetilde{\vy}_i = \mU_i \vy_i,
            \qquad
            \widetilde{\vx}_i \defeq \mA^\top\widetilde{\vy}_i,
            \qquad
            \vx_i \defeq \mV_{i-1}^\top \widetilde{\vx}_i
            \quad\text{and}\quad
            \mV_i = \mV_{i-1}.
        \]
\end{itemize}
    Construct \(\cT_i\) this way until the algorithm halts or uses at least \(d\) queries.
If it halts after computing \(k \leq d-1\) queries, let \(\vz\in\bbR^n\) be the vector that \(\Phi\) returns on the transcript \(\cT_k\).
    Let \(\mU \defeq \mU_k\) and \(\mV \geqq_{k+1} \mV_{k}\) on domain \(\{\vx_1,\ldots,\vx_k,\vz\}\) and range \(\{\vr_1,\ldots,\vr_{k+1}\}\).
    By the nested construction of our rotations, we can verify that \(\mU\vb=\vb\) and that
    \[
    \begin{cases}
        \vy_i = \mU^\top\mA\mV\vx_i & \text{where } \vx_i = \Phi_i(\cT_{i-1}) \text{ for } i \text{ odd, and} \\
        \vx_i = (\mU^\top\mA\mV)^\top\vy_i & \text{where } \vy_i = \Phi_i(\cT_{i-1}) \text{ for } i \text{ even}.
    \end{cases}
    \]
    Therefore, there exists \(\mB=\mU^\top\mA\mV\) such that \(\cT_k\) is the transcript returned by \(\Phi(\mB,\vb)\).
    Further, since \(\Span\{\vr_1,\ldots,\vr_{k+1}\} = \cR_{k}(\mA,\vb)\), we know that \(\mV\vz\in\cR_{k}(\mA,\vb)\).
    This completes the first case.

    In the other case, \(\Phi\) uses at least \(d\) queries.
    Then let \(\mU \defeq \mU_{d-1}\) and \(\mV \defeq \mV_{d-1}\).
    Thereby, we know that \(\Phi(\mU^\top\mA\mV,\vb)\) uses at least \(d\) queries, completing the proof.
\end{proof}

Note that the returned vector \(\vz\) is related to the right Krylov subspace \(\cR_{k}(\mA,\vb)\), which suggests that a Krylov method can be used to recover the performance of \(\Phi\).
However, keep in mind that it takes \(2k+1\) matvecs with \mA or \(\mA^\top\) to construct \(\cR_{k}(\mA,\vb) = \cK_{k}(\mA^\top\mA,\mA^\top\vb)\).
Our lower bound \cref{thm:krylov-opt-alternating} has a factor of 2 in it because of this doubling of \(k\) needed to construct \(\cR_k(\mA,\vb)\).
That said, we are now ready to prove \cref{thm:krylov-opt-alternating}.

\begin{proof}[Proof of \cref{thm:krylov-opt-alternating}]
    Fix any matrix \(\mA\in\cF\) and apply \cref{thm:krylov-det-simulate} to construct the matrix \(\mB=\mU^\top\mA\mV\).
    
    Let \(d\) be the grade of \mA with respect to \vb.
    For any invertible matrix \mA and vector \vb, we have \(\mA^{-1}\vb \in \cR_{d-1}(\mA,\vb)\).
    Therefore, forming the Krylov subspace for the normal equations computes \(2d-1\) matvecs in the worst case.
    So, if \(t_{\Phi}(\mB,\vb) \geq d\), then we have \(t_{\rm k\text{-}opt}(\mA,\vb) \leq 2d-1 \leq 2t_{\Phi}(\mB,\vb)-1\).

    Otherwise \(\Phi(\mB,\vb)\) returns a vector \(\vz\) such that \(\textsc{acc}(\mB,\vb,\vz)=1\) after \(t_{\Phi}(\mB,\vb) = k < d\) matvecs.
    By the orthogonal invariance of our test, we know that \(\textsc{acc}(\mA,\vb,\mV\vz)=1\).
    Since \(\mV\vz\in\cR_{k}(\mA,\vb)\), we know that \(t_{\rm k\text{-}opt}(\mA,\vb) \leq 2k+1 = 2t_{\Phi}(\mB,\vb)+1\).

    Between these two cases, we have ensured that for all \(\mA\in\cF\) there exists a matrix \(\mB\in\cF\) such that
    \[
        t_{\rm k\text{-}opt}(\mA,\vb) \leq 2t_{\Phi}(\mB,\vb) +1 \leq 2 \sup_{\mB\in\cF}t_{\Phi}(\mB,\vb) +1.
    \]
    Taking the supremum over all \(\mA\in\cF\) completes the proof.
\end{proof}

The proof of \cref{thm:krylov-opt}, which gives a factor of 4 over all (possibly non-alternating) methods, follows from \cref{thm:krylov-opt-alternating} by observing that alternating matvec algorithms can simulate non-alternating ones by (at most) doubling their matvec complexity.
The proof of the main theorem \cref{thm:det-lower-bound-to-krylov} follows from \cref{thm:krylov-opt} by taking \(\textsc{acc}\) to be the backward error test shown in \cref{eq:acc-berr}.

\section{Inapproximability by odd polynomials}\label{sec:odd-poly-inapprox}

In this section, we prove \cref{thm:odd-poly-inapprox}, showing that no odd polynomial can minimize the backward error as characterized by a Krylov method run on the normal equations without depending on the condition number.
\begin{reptheorem}{thm:odd-poly-inapprox}[Hardness of approximation with odd polynomials]
    Fix \(n \gtrsim 1\), \(\kappa \gtrsim 1\), and \(\kappa^{-0.1} \lesssim \eps \lesssim 1\).
    Then there exist weights \(w_1,\ldots,w_n\in[0,1]\) and nodes \(\sigma_1,\ldots,\sigma_n\in[1/\kappa,1]\) such that every odd polynomial \(p\) with
    \begin{equation}
        \label{eq:odd-weighted-l2-feasibility}
        \sum_{i=1}^n w_i (\sigma_i p(\sigma_i) - 1)^2
        \leq \sigma_1^2 \eps^2 \sum_{i=1}^n w_i p(\sigma_i)^2
    \end{equation}
    has degree at least \(t\gtrsim \min\{n, \log(\kappa)/\eps\}\).
\end{reptheorem}
We will prove this in two stages.
First, we prove a continuous version of this hardness result, essentially giving the proof when \(n=\infty\).
Then, a basic application of Gaussian quadrature is used to recover \cref{thm:odd-poly-inapprox}.

\subsection{A lower bound with a measure}

In this section, we prove a weaker form of \cref{thm:odd-poly-inapprox}, left in terms of a measure.
For any measure \(\mu\), write \(\norm{f}_{L_2(\mu)}^2 \defeq \int |f(x)|^2 d\mu(x)\) as the \(L_2\) norm with respect to \(\mu\).
We will focus on the measure that puts about half of its mass on \(x_0 \defeq 1/\kappa\), and the rest on the continuous interval \(\cI_\eps \defeq [2\eps,1]\):
\begin{equation}
  \label{eq:odd-weighted-l2-hard-dist}
  \mu \defeq \delta_{\kappa^{-1}} + (1-2\eps)\textsc{Unif}[2\eps,1].
\end{equation}
For this measure, we can prove the following lower bound.
\begin{theorem}[Hardness of weighted approximation, continuous version]
  \label{thm:odd-weighted-l2-lower-bound-cts}
  Fix \(\kappa \gtrsim 1\), and \(\kappa^{-0.1} \lesssim \eps \lesssim 1\).
  Any odd polynomial \(p\) that has
  \begin{equation}
      \label{eq:cts-feasibility}
      \norm{xp - 1}_{L_2(\mu)}^2 \leq \eps^2 \norm{p}_{L_2(\mu)}^2
  \end{equation}
  must have degree \(t \gtrsim \log(\kappa)/\eps\).
\end{theorem}

To prove this fact, we will need several standard bounds about polynomials.
The first two bounds assert that polynomials cannot be too spikey: Bernstein's inequality bounds the derivative of \(p\) and the pointwise Remez inequality essentially bounds the value of a polynomial on \([-\eps,\eps]\) based on the value of the polynomial on \([-1,-\eps]\cup[\eps,1]\).
The third bound lets us transfer between the \(L_2\) norm and the sup norm.

\begin{fact}[Polynomial bounds]
  Let \(\norm{f}_{\cS} \defeq \sup_{x\in\cS}|f(x)|\) and \(\norm{f}_{L_2(\cS)}^2 \defeq \int_{\cS} |f(x)|^2 dx\).
  For a polynomial \(p\) of degree at most \(t\),
  \begin{enumerate}
    \item \textbf{Bernstein}\textbf{.}~ \(|p'(x)| \lesssim t \norm{p}_{[-1,1]}\) for all \(|x| \leq 1/2\).
    \item \textbf{Pointwise Remez}\textbf.~ \(\norm{p}_{[-1,1]} \lesssim e^{\rC t \eps} \norm{p}_{[-1,-\eps]\cup[\eps,1]}\) for any \(0 \leq \eps \leq 1/2\).
    \item \textbf{Nikolskii}\textbf.~ \(\norm{p}_{\cI} \lesssim t \norm{p}_{L_2(\cI)}\) for any interval \cI of width \(|\cI| \geq 3/4\).
  \end{enumerate}
  These are consequences of \cite[Thm.~5.1.7]{borwein2012polynomials}, \cite[Exc.~A.4.E.21]{borwein2012polynomials}, and  \cite[Thm.~A.4.4]{borwein2012polynomials}, respectively.
\end{fact}

\noindent
Following our construction of \(\mu\), the feasibility condition (\cref{eq:cts-feasibility}) is equivalent to
\begin{equation}
  \label{eq:weighted-l2-feasibility-cts}
  (x_0 p(x_0)-1)^2 + \norm{xp - 1}_{L_2(\cI_\eps)}^2 \leq \eps^2 p(x_0)^2 + \eps^2 \norm{p}_{L_2(\cI_\eps)}^2.
\end{equation}
The proof of \cref{thm:odd-weighted-l2-lower-bound-cts} uses two properties about any polynomial satisfying the feasibility condition (\cref{eq:weighted-l2-feasibility-cts}): the fact that \(|p(x_0)| \gtrsim 1/\eps\) and the fact that \(\norm{p}_{L_2(\cI_\eps)} \lesssim |p(x_0)|\).
\cref{lem:weighted-l2-large-atom} proves the first fact, and \cref{lem:weighted-l2-small-tail} proves the second.
\begin{lemma}[\(p\) is large at \(x_0\)]
  \label{lem:weighted-l2-large-atom}
  Let \(\kappa^{-0.1} \leq \eps \lesssim 1\).
  Let \(p\) be any polynomial satisfying \cref{eq:weighted-l2-feasibility-cts}, and recall that \(x_0 = \kappa^{-1}\).
  Then \(|p(x_0)| \gtrsim 1/\eps\).
\end{lemma}
\begin{proof}
  Fix any odd polynomial \(p\) that satisfies \cref{eq:weighted-l2-feasibility-cts} and let \(P \defeq |p(x_0)|\).
  For any \(a > 0\), we have that the parabola \(ap^2 + bp + c \geq c - \frac{b^2}{4a}\) for all \(p\in\bbR\).
  Thereby, we can bound
  \[
    (xp(x)-1)^2 - \eps^2 p(x)^2 = (x^2 - \eps^2)p(x)^2 -2xp(x) + 1
    \geq 1 - \tfrac{x^2}{x^2 - \eps^2}
    = -\tfrac{\eps^2}{x^2 - \eps^2},
  \]
  which is at least \(-2\eps^2/x^2\) for \(x\geq2\eps\).
  Integrating over all \(x\in\cI_\eps\), we get
  \[
    \norm{xp - 1}_{L_2(\cI_\eps)}^2 - \eps^2\norm{p}_{L_2(\cI_\eps)}^2 \geq -2\eps^2 \int_{2\eps}^1 x^{-2} dx \geq -\eps.
  \]
  Separately, recalling that \(x_0 = 1/\kappa \leq \eps\), we can bound
  \[
    \eps^2 P^2 - (x_0 P - 1)^2 
    = (\eps^2 - x_0^2)P^2 + 2x_0 P - 1
    \leq \eps^2 P^2 + 2\eps P - 1
    = (\eps P + 1)^2 - 2.
  \]
  Combining our above facts with a rearrangement of the feasibility equation (\cref{eq:weighted-l2-feasibility-cts}) gives us
  \begin{align*}
    (\eps P + 1)^2 - 2
    \geq \eps^2 P^2 - (x_0 P - 1)^2
    \geq \norm{xp - 1}_{L_2(\cI_\eps)}^2 - \eps^2\norm{p}_{L_2(\cI_\eps)}^2
    \geq -\eps.
  \end{align*}
  Solving this inequality for \(P\) completes the proof.
\end{proof}

\begin{lemma}[The tail mass of \(p\) is bounded by \(p(x_0)\)]
  \label{lem:weighted-l2-small-tail}
  Let \(\kappa^{-0.1} \leq \eps \lesssim 1\).
  Let \(p\) be any polynomial satisfying \cref{eq:weighted-l2-feasibility-cts}, and recall that \(x_0 = \kappa^{-1}\).
  Then \(\norm{p}_{L_2(\cI_\eps)} \lesssim |p(x_0)|\).
\end{lemma}
\begin{proof}
  Using that \(x \geq 2\eps\) for \(x\in\cI_\eps\), and the generalized triangle inequality, we have that
  \[
    2\eps^2\norm{p}_{L_2(\cI_\eps)}^2
    \leq \frac12 \norm{xp}_{L_2(\cI_\eps)}^2
    \leq \norm{xp-1}_{L_2(\cI_\eps)}^2 + 1.
  \]
  Rearranging the feasibility equation tells us that
  \[
    \norm{x p - 1}_{L_2(\cI_\eps)}^2 - \eps^2\norm{p}_{L_2(\cI_\eps)}^2
    \leq \eps^2 p(x_0)^2 - (x_0 p(x_0) - 1)^2
    \leq \eps^2 p(x_0)^2.
  \]
  Combining the above inequalities gives
  \[
    \norm{p}_{L_2(\cI_\eps)}^2 \leq p(x_0)^2 + \frac1{\eps^2}.
  \]
  Substituting in the bound \(\frac1{\eps^2} \lesssim p(x_0)^2\) from \cref{lem:weighted-l2-large-atom} completes the proof.
\end{proof}

We are now ready to prove the continuous lower bound.
\begin{proof}[Proof of \cref{thm:odd-weighted-l2-lower-bound-cts}]
  Let \(p\) be any odd polynomial that satisfies \cref{eq:weighted-l2-feasibility-cts}.
Since \(p(0)=0\), the mean value theorem guarantees there exists some \(x_\star \in [0,1/\kappa]\) such that \(|p'(x_\star)| \gtrsim \kappa |p(x_0)|\).
  By chaining together Bernstein, Remez, Nikolskii, the fact that \(p\) is odd, and \cref{lem:weighted-l2-small-tail} we find that
  \[
    \kappa |p(x_0)|
    \lesssim |p'(x_\star)|
    \lesssim t \norm{p}_{[-1,1]}
    \lesssim t e^{\rC t \eps} \norm{p}_{[-1,-2\eps]\cup[2\eps,1]}
    = t e^{\rC t \eps} \norm{p}_{[2\eps,1]}
    \lesssim t^2 e^{\rC t \eps} \norm{p}_{L_2([2\eps,1])}
    \lesssim t^2 e^{\rC t \eps} |p(x_0)|.
\]
  Equivalently, we have
  \[
    t \gtrsim \frac{\log(\kappa)}{\eps} - \frac{\log(t)}{\eps}.
  \]
  For \(\eps > \kappa^{-0.1}\), the above inequality implies that \(t \gtrsim \log(\kappa)/\eps\), completing the proof.
\end{proof}

\subsection{Making the problem finite}

To transform \cref{thm:odd-weighted-l2-lower-bound-cts} into the finite-support result \cref{thm:odd-poly-inapprox}, we make use of the classic Gaussian quadrature observation:
\begin{importedtheorem}[Positive Gaussian Quadrature \protect{\cite[Thm.~3.5]{chen2024lanczos}}]
  Let \(\mu\) be a (Borel) measure supported on a bounded interval \cI, and suppose that \cI contains at least \(n\) distinct points.
  Then, there exist nodes \(\sigma_1,\ldots,\sigma_n\in\cI\) and weights \(w_1,\ldots,w_n\in[0,1]\) such that
  \[
    \int p(x) d\mu(x) = \sum_{i=1}^n w_i p(\sigma_i)
    \qquad
    \text{for all }
    \deg(p) \leq 2n-1.
  \]
\end{importedtheorem}
\begin{proof}[Proof of \cref{thm:odd-poly-inapprox}]
Take \(\sigma_1 \geq \ldots \geq \sigma_n\) and \(w_1,\ldots,w_n\) to be the Gaussian quadrature of our distribution \(\mu\).
Let \(p\) be an odd polynomial that satisfies the feasibility criteria \cref{eq:odd-weighted-l2-feasibility}.
Since \(\sigma_1 \leq 1\) we know that
\[
  \sum_{i=1}^n w_i (\sigma_i p(\sigma_i)-1)^2 \leq \eps^2 \sum_{i=1}^n w_i p(\sigma_i)^2.
\]
Since \(x \mapsto (x p(x) - 1)^2\) and \(x \mapsto p(x)^2\) are polynomials of degree at most \(2t+2\), we know that if \(p\) has degree \(t \lesssim n\) then we have the continuous bound \(\norm{xp-1}_{L_2(\mu)}^2 \leq \eps^2 \norm{p}_{L_2(\mu)}^2\).
Therefore \(p\) must have degree \(t \gtrsim \log(\kappa)/\eps\).
\end{proof}

\section{Upper bound for symmetric matrices}
\label{sec:minberr-symmetric-upper}

In this section, we prove \cref{thm:indefinite-poly-approx}, which guarantees that MINBERR converges in at most \(5/\eps\) matvecs on symmetric indefinite matrices.
We restate the theorem below for convenience.

\begin{reptheorem}{thm:indefinite-poly-approx}[Upper bound for symmetric MINBERR approximation]
Fix \(\eps>0\) and probability distribution \(\mu\) supported on \([-1,1]\).
  Then, there exists a polynomial \(p\) of degree at most \(k=\lceil 5/\eps \rceil\) such that
  \begin{equation}
    \label{eq:symmetric-poly-upper}
    \norm{xp - 1}_{L_2(\mu)} \leq \eps \norm{p}_{L_2(\mu)}.
  \end{equation}
\end{reptheorem}

The proof of this result works by case analysis, depending on the measure \(\mu\).
If \(\mu\) does not place a lot of mass near zero, then we may use a typical approximation to the inverse function, such as the one below.
\begin{lemma}[Relative error approximation to the inverse]
  \label{lem:rel-error-inverse-poly}
  Fix integer \(k\geq1\) and let \(\tau = \sqrt8/k\).
  There exists a polynomial \(p(x)\) of degree at most \(k\) with
  \[
    |xp(x)-1| \leq 1
    \quad\text{and}\quad
    |p(x) - x^{-1}| \leq \frac{\tau^2}{2|x|^3}
    \quad
    \text{for all } x\in[-1,1].
  \]
\end{lemma}
If \(\mu\) does place a lot of mass near zero, we essentially ignore the task of trying to approximate the inverse function and instead pick a bump polynomial that is concentrated near zero.
Such functions will have \(\norm{xp}_{L_2(\mu)} \ll \norm{p}_{L_2(\mu)}\), since \(x\) is small where the mass of \(\mu\) is concentrated.
We pick the following bump function in an attempt to slightly optimize constants.
\begin{lemma}[Polynomial bump function]
    \label{lem:bump-inverse-poly}
    Fix integer \(k\geq1\) and let \(\tau = \sqrt8/k\).
    There exists a polynomial \(f(x)\) of degree at most \(k\) such that
    \[
        0 \leq f(x) \leq \frac{\tau^2}{2 x^2} \text{ for all } |x| \leq 1,
        \quad\text{and}\quad
        \bigl(f(x)\bigr)^2 \geq \frac{4}{17} \text{ for all } |x| \leq \tau.
    \]
\end{lemma}
\begin{proof}[Proof of \cref{thm:indefinite-poly-approx}]
    We start by forming the test to determine if \(\mu\) has a lot of mass near zero.
    To do so, define the threshold \(\tau = \sqrt 8 / k\) and define the function \(g(x) = x^{-1} \mathbbm1_{\{|x|>\tau\}}\), so that \(\norm{xg-1}_{L_2(\mu)}^2 = \int_{|x|\leq\tau}d\mu(x)\).
We proceed casewise depending on if the following condition holds:
    \begin{equation}
        \label{eq:g-test}
        \norm{xg-1}_{L_2(\mu)}^2 \leq \frac{17}{4k^2} \norm{g}_{L_2(\mu)}^2.
    \end{equation}
    \textit{Case 1: \cref{eq:g-test} holds.~}
    We use the approximate inverse \(p(x)\) from \cref{lem:rel-error-inverse-poly}, which has
    \[
        |p(x)| \geq \frac1{2|x|}
        \quad\text{and}\quad
        |xp(x)-1| = |x| |p(x)-x^{-1}| \leq \frac{\tau^2}{2x^2}
        \quad\text{for } |x| \geq \tau.
    \]
    This implies that \(\norm{p}_{L_2(\mu)}^2 \geq \frac14\norm{g}_{L_2(\mu)}^2\).
    Further, using the bound \(|xp(x)-1|\leq1\) for \(|x|\leq\tau\), we have
    \[
        \norm{xp-1}_{L_2(\mu)}^2
\leq \!\int\limits_{|x|\leq\tau}\!\!d\mu(x) + \!\int\limits_{|x|>\tau} \!\!\tfrac{\tau^4}{4x^4}d\mu(x)
        \leq \norm{xg-1}_{L_2(\mu)}^2 + \!\int\limits_{|x|>\tau} \!\!\tfrac{\tau^2}{4x^2}d\mu(x)
        \leq \norm{xg-1}_{L_2(\mu)}^2 + \tfrac{2}{k^2}\norm{g}_{L_2(\mu)}^2,
    \]
    which is at most \(\frac{25}{4k^2}\norm{g}_{L_2(\mu)}^2\leq\frac{25}{k^2}\norm{p}_{L_2(\mu)}^2\), completing this case.\\

    \noindent
    \textit{Case 2: \cref{eq:g-test} does not hold.~}
    Let \(f(x)\) be the bump polynomial from \cref{lem:bump-inverse-poly}.
    We have that
    \[
        \norm{f}_{L_2(\mu)}^2
        \geq \frac{4}{17}\int\limits_{|x|\leq\tau}d\mu(x)
        = \frac4{17}\norm{xg-1}_{L_2(\mu)}^2
        \geq \frac1{k^2}\norm{g}_{L_2(\mu)}^2.
    \]
    Now, we bound the residual-like term
    \[
        \norm{xf}_{L_2(\mu)}^2
        \leq \tau^2 \!\int\limits_{|x|\leq\tau}\!\!|f(x)|^2 d\mu(x) + \!\int\limits_{|x|>\tau}\!\!\tfrac{\tau^4}{4x^2}d\mu(x)
        \leq \frac{8}{k^2}\norm{f}_{L_2(\mu)}^2 + \frac{16}{k^4}\norm{g}_{L_2(\mu)}^2
        \leq \frac{24}{k^2}\norm{f}_{L_2(\mu)}^2.
    \]
    We then define the polynomial \(\widehat p(x) \defeq \rC f(x)\), and examine the limit
    \[
        \lim_{\rC \to \infty} \frac{\norm{x\widehat p - 1}_{L_2(\mu)}}{\norm{\widehat p}_{L_2(\mu)}}
        = 
        \lim_{\rC \to \infty} \frac{\norm{xf - \rC^{-1}}_{L_2(\mu)}}{\norm{f}_{L_2(\mu)}}
        = \frac{\norm{xf}_{L_2(\mu)}}{\norm{f}_{L_2(\mu)}}
        \leq \frac{\sqrt{24}}{k}.
    \]
    So, some finite \(\rC\) yields \(\norm{x\widehat p - 1}_{L_2(\mu)} \leq (5/k) \norm{\widehat p}_{L_2(\mu)}\), completing this case, and thereby the proof.
\end{proof}

\subsection{Constructing the polynomials}

We now show how to construct the polynomials in \cref{lem:rel-error-inverse-poly,lem:bump-inverse-poly}.
Both of these constructions are slight twists and scalings of the Chebyshev polynomial of the second kind, nudged in part just to optimize constants.
These constructions are also closely related to \emph{Fej\'er polynomials}, and none of the analysis in this subsection is likely novel at all.
More precisely, \cite[Pg.~153]{dzyadyk2008theory} asserts that the following function is a degree \(2m-2\) polynomial:
\[
    p(x) = \left(\frac{\sin(m\theta)}{\sin(\theta)}\right)^2
    \qquad\text{where } \theta = \arcsin(x).
\]

\begin{proof}[Proof of \cref{lem:rel-error-inverse-poly}]
    Let \(m = \lceil k/2\rceil\) and define \(\alpha = k/2m \in (0,1]\).
    Then invoke the polynomial
    \[
        r(x) = \left(\frac{\sin(m\theta)}{m\sin\theta}\right)^2
        \qquad\text{where } \theta = \arcsin(\alpha x).
    \]
    We will then define \(p(x) = (1-r(x))/x\).
    First, since \(r(0)=1\), we have that \(p(x)\) is a polynomial.
    Second, since \(|\sin(m\theta)| \leq m|\sin\theta|\) and \(|\sin(m\theta)|\leq1\) for all integers \(m \geq 1\), we have
    \[
        0 \leq r(x) \leq 1
        \quad
        \text{and}
        \quad
        r(x) \leq \frac{1}{m^2\alpha^2x^2} = \frac{\tau^2}{2x^2}.
    \]
    Since \(r(x)=xp(x)-1\), the first inequality gives \(|xp(x)-1|\leq1\) for all \(x\in[-1,1]\).
    The second bound gives \(|p(x) - x^{-1}| = |r(x)/x| \leq \tau^2/(2|x|^3)\), completing the proof.
\end{proof}
\begin{proof}[Proof of \cref{lem:bump-inverse-poly}]
    Let \(m = \lceil k/2\rceil\) and define \(\alpha = k/2m\in(0,1]\).
    Define the polynomial
    \[
        f(x) = \left(\frac{\sin(m\theta)}{m\sin(\theta)}\right)^2
        \qquad\text{where } \theta = \arcsin(\alpha x).
    \]
    When \(m=1\), we have \(f(x)=1\), which trivially satisfies the guarantee.
    So, we assume \(m\geq2\).
    First note that \(0 \leq f(x) \leq \tau^2/2x^2\), as shown in the proof of \cref{lem:rel-error-inverse-poly}.
    The bound on \((f(x))^2\) is more involved.
    Since \(f(x)\) is symmetric, it suffices to lower bound \((f(x))^2\) only for \(x\in[0,\tau]\).
    Since \(z\leq\arcsin(z)\leq\frac{\pi}{\sqrt 8}z\) for \(z\in[0,1/\sqrt 2]\), we have
    \[
    0 \leq m\alpha x \leq m \theta \leq \frac{\pi}{\sqrt 8} m\alpha x
    \leq \frac{\pi}{\sqrt 8} m\alpha\tau \leq \frac{\pi}{2}.
    \]
    Further, we directly have \(m\alpha x \leq m \alpha\tau = \sqrt 2\).
    Using that \(\sin(z)\) is increasing and \(\sin(z)/z\) is decreasing on \([0,\pi/2]\), we conclude
    \[
        (f(x))^2
        = \left(\frac{\sin(m\theta)}{m\alpha x}\right)^4
        \geq \left(\frac{\sin(m\alpha x)}{m\alpha x}\right)^4
        \geq \left(\frac{\sin(\sqrt2)}{\sqrt 2}\right)^4
        \geq \frac{4}{17}.
    \]
\end{proof}

\section{Lower bounds for randomized methods on symmetric matrices}
\label{sec:symmetric-lower-bounds}

In this section we prove the $\Omega(1/\eps)$ lower bound on the number of matvecs needed for even a randomized algorithm to solve a  symmetric linear system to small backward error, as stated in \cref{thm:main-minberr-symmetric}. Our argument can also be used to show a $\Omega(1/\sqrt\eps)$ lower bound for psd linear systems.
To begin, we prove \cref{lem:bwd-to-res-reduction}, which shows that any algorithm which achieves \(\eps\) backward error can be used to achieve \(\cO(1)\) residual on any matrix of condition number at most \(1/\eps\).

This result is an elementary consequence of a well known fact: for a system of condition number \(\kappa\) and for every vector \vx, the relative residual of \vx is at most \(\kappa\) times the backward error of \vx.
For the purpose of optimizing constants, we state a small strengthening of this fact.
\begin{replemma}{lem:bwd-to-res-reduction}[Backward-to-residual reduction]
  Let \(\eps > 0\) and let \mA have condition number at most \(\kappa\).
  Fix any vectors \vb and \(\widetilde\vx\) such that \(\mathrm{berr}(\mA,\vb,\widetilde\vx) \leq \eps\).
  Then, using a single matvec with \mA, we can construct a vector \(\widehat\vx\) such that \(\norm{\mA\widehat\vx-\vb}_2\leq\kappa\eps\norm{\vb}_2\).
\end{replemma}
\begin{proof}
Let \(\widehat\vx\) be the vector such that \(\mA\widehat\vx\) is the projection of \(\vb\) onto \(\Span\{\mA\widetilde\vx\}\).
  Namely, \(\mA\widetilde\vx\) is orthogonal to \(\mA\widehat\vx-\vb\).
  We can construct \(\widehat\vx\) with one matvec, returning \(\widehat\vx = (\vz^\top\vb/\vz^\top\vz) \widetilde\vx\) where \(\vz \defeq \mA\widetilde\vx\).
Now recall that for any triangle, area is equal to half of the base times the height and is upper bounded by half the product of any two sides.
  So, for the triangle with vertices \(\vec0\), \vb, and \(\mA\widetilde\vx\) we have
  \[
    \norm{\mA\widetilde\vx}_2 \norm{\mA\widehat\vx-\vb}_2
    = 2 \cdot \operatorname{area}
    \leq \norm{\vb}_2 \norm{\mA\widetilde\vx-\vb}_2.
  \]
  Rearranging this formula, we get that the relative residual of \(\widehat\vx\) is at most
  \[
    \frac{\norm{\mA\widehat\vx-\vb}_2}{\norm{\vb}_2}
    \leq \frac{\norm{\mA\widetilde\vx-\vb}_2}{\norm{\mA\widetilde\vx}_2}
    \leq \frac{\eps \norm{\mA}_2\norm{\widetilde\vx}_2}{\norm{\mA\widetilde\vx}_2}
    \leq \eps\kappa.
  \]
\end{proof}

The crucial idea is to begin with a method that solves \(\mA\vx=\vb\) to backward error \(\eps\).
If we apply this algorithm on a matrix with condition number \(\kappa < 1/\eps\), then the resulting solution has relative residual \(\cO(1)\).
We then appeal to a recent line of work \cite{chewi2024query,derezinski2026matrix,shamir2026simple} which settled the complexity of solving a linear system of condition number \(\kappa\) to relative residual \(\eta\).
\begin{importedtheorem}[Relative residual matvec lower bounds \protect{\cite{shamir2026simple}}]
  \label{thm:shamir-lower-bound}
  Let \(\kappa \geq 9\) and \(0 < \eta < 1\).
  Consider a (possibly randomized) algorithm that is given a vector \vb, computes \(t\) matvecs with any symmetric matrix \mA of condition number at most \(\kappa\), and returns a vector \(\widetilde\vx\) with relative residual \(\eta\):
  \[
    \norm{\mA\widetilde\vx-\vb}_2 \leq \eta \norm{\vb}_2 \quad\text{with prob. } \geq 2/3.
  \]
  Then, this algorithm must compute at least \(t \geq (1 - o(1)) \kappa\log(1/\eta)\) matvecs in the worst case, where the little-oh shrinks as \(\kappa\) grows large.
  If we further assume that \mA is psd, the lower bound tightens to \(t \geq (1-o(1)) \frac12 \sqrt\kappa\log(1/\eta)\).
\end{importedtheorem}

We can immediately prove the lower bounds on backward error matvec complexity, completing the proof of \cref{thm:main-minberr-symmetric}.
\begin{theorem}[Symmetric and psd lower bounds for backward error]
  \label{thm:symmetric-psd-lower-bound}
  Every (possibly randomized and possibly adaptive) matvec algorithm which can solve any symmetric linear system to backward error at most \(\eps\) with probability at least \(2/3\) must use at least \(t \geq (e^{-1}-o(1)) / \eps\) matvecs, where the little-oh shrinks as \(\eps\) grows small.
  For psd systems, the rate shrinks to \(t \geq (e^{-1}-o(1))/\sqrt{\eps}\) matvecs.
\end{theorem}
\begin{proof}
  Consider the set of symmetric matrices \(\cA = \{\mA ~:~ \cond(\mA) \leq \kappa\}\) where \(\kappa \defeq \alpha/\eps\) for some \(\alpha\in(0,1)\).  
  Any method which achieves backward error \(\eps\) on all symmetric matrices, by \cref{lem:bwd-to-res-reduction}, therefore achieves relative residual at most \(\eta=\alpha\).
  Optimizing over \(\alpha\) gives that \(\kappa\log(1/\eta) = 1/e\eps\).
  Hence, \cref{thm:shamir-lower-bound} tells us that achieving this relative residual over the set \cA requires \(t \geq (e^{-1}-o(1)) / \eps\) matvecs.
  The same argument extends directly to the psd case, where we maximize \(\frac12\sqrt\alpha\log(1/\alpha)\) by picking \(\alpha=e^{-2}\), which yields the advertised bound.
\end{proof}
 
\subsection*{AI Disclosure}
We used various ChatGPT 5.6, 6.0, and 6.1 models, as well as Gemini 4 Argon to assist with proving the theorems in this article.
Human-AI interaction was used to simplify and streamline the analysis.
The authors verified the correctness and originality of all content, including references.

\subsection*{Acknowledgments}
We thank Ethan Epperly for helpful conversations.
MWM and RAM would like to acknowledge the DARPA DIAL and DARPA AIQ programs for providing partial support of this work. MD was supported in part by NSF CAREER Grant CCF-233865 and a Google ML and Systems Junior Faculty Award. CM was supported by NSF  Grant CAIF awards 2615856 and 2427363.
RAM's research was supported in part by a grant of access to OpenAI models through the ChatGPT for Academic Researchers program.

\bibliographystyle{halpha}
{\footnotesize\newcommand{\etalchar}[1]{$^{#1}$}

 }

\end{document}